\documentclass[10pt]{article}
\usepackage{amsmath,amssymb}
\usepackage{newtxtext,newtxmath,bm}
\usepackage{graphics,graphicx,epsfig}
\usepackage[square,numbers,sort&compress]{natbib}
\usepackage{tabularx,array,multirow,multicol,booktabs}
\usepackage{algorithm,algpseudocode}
\usepackage[hmargin=1in,vmargin=1in]{geometry}
\numberwithin{equation}{section}

\begin{document}
%
%
\newcommand{\calA}{\mathcal{A}}%
\newcommand{\calB}{\mathcal{B}}%
\newcommand{\calC}{\mathcal{C}}%
\newcommand{\calD}{\mathcal{D}}%
\newcommand{\calE}{\mathcal{E}}%
\newcommand{\calF}{\mathcal{F}}%
\newcommand{\calH}{\mathcal{H}}%
\newcommand{\calI}{\mathcal{I}}%
\newcommand{\calJ}{\mathcal{J}}%
\newcommand{\calK}{\mathcal{K}}%
\newcommand{\calL}{\mathcal{L}}%
\newcommand{\calM}{\mathcal{M}}%
\newcommand{\calN}{\mathcal{N}}%
\newcommand{\calP}{\mathcal{P}}%
\newcommand{\calQ}{\mathcal{Q}}%
\newcommand{\calR}{\mathcal{R}}%
\newcommand{\calS}{\mathcal{S}}%
\newcommand{\calV}{\mathcal{V}}%
\newcommand{\calW}{\mathcal{W}}%
\newcommand{\calX}{\mathcal{X}}%
\newcommand{\calT}{\mathcal{T}}%
\newcommand{\bfa}{{\bf a}}\newcommand{\bfA}{{\bf A}}%
\newcommand{\bfb}{{\bf b}}\newcommand{\bfB}{{\bf B}}%
\newcommand{\bfc}{{\bf c}}\newcommand{\bfC}{{\bf C}}%
\newcommand{\bfd}{{\bf d}}\newcommand{\bfD}{{\bf D}}%
\newcommand{\bfe}{{\bf e}}\newcommand{\bfE}{{\bf E}}%
\newcommand{\bff}{{\bf f}}\newcommand{\bfF}{{\bf F}}%
\newcommand{\bfg}{{\bf g}}\newcommand{\bfG}{{\bf G}}%
\newcommand{\bfh}{{\bf h}}\newcommand{\bfH}{{\bf H}}%
\newcommand{\bfi}{{\bf i}}\newcommand{\bfI}{{\bf I}}%
\newcommand{\bfj}{{\bf j}}\newcommand{\bfJ}{{\bf J}}%
\newcommand{\bfk}{{\bf k}}\newcommand{\bfK}{{\bf K}}%
\newcommand{\bfl}{{\bf l}}\newcommand{\bfL}{{\bf L}}%
\newcommand{\bfm}{{\bf m}}\newcommand{\bfM}{{\bf M}}%
\newcommand{\bfn}{{\bf n}}\newcommand{\bfN}{{\bf N}}%
\newcommand{\bfo}{{\bf o}}\newcommand{\bfO}{{\bf O}}%
\newcommand{\bfp}{{\bf p}}\newcommand{\bfP}{{\bf P}}%
\newcommand{\bfq}{{\bf q}}\newcommand{\bfQ}{{\bf Q}}%
\newcommand{\bfr}{{\bf r}}\newcommand{\bfR}{{\bf R}}%
\newcommand{\bfs}{{\bf s}}\newcommand{\bfS}{{\bf S}}%
\newcommand{\bft}{{\bf t}}\newcommand{\bfT}{{\bf T}}%
\newcommand{\bfu}{{\bf u}}\newcommand{\bfU}{{\bf U}}%
\newcommand{\bfv}{{\bf v}}\newcommand{\bfV}{{\bf V}}%
\newcommand{\bfw}{{\bf w}}\newcommand{\bfW}{{\bf W}}%
\newcommand{\bfx}{{\bf x}}\newcommand{\bfX}{{\bf X}}%
\newcommand{\bfy}{{\bf y}}\newcommand{\bfY}{{\bf Y}}%
\newcommand{\bfz}{{\bf z}}\newcommand{\bfZ}{{\bf Z}}%
\newcommand{\bfalpha}{\bm{\alpha}}%
\newcommand{\bfbeta}{\bm{\beta}}%
\newcommand{\bfgamma}{\bm{\gamma}}%
\newcommand{\bfGamma}{\boldsymbol{\Gamma}}%
\newcommand{\bfdelta}{\bm{\delta}}%
\newcommand{\bfepsilon}{\bm{\epsilon}}%
\newcommand{\bfzeta}{\bm{\zeta}}%
\newcommand{\bfeta}{\bm{\eta}}%
\newcommand{\bftheta}{\bm{\theta}}%
\newcommand{\bfiota}{\bm{\iota}}%
\newcommand{\bfkappa}{\bm{\kappa}}%
\newcommand{\bflambda}{\bm{\lambda}}%
\newcommand{\bfLambda}{\boldsymbol{\Lambda}}%
\newcommand{\bfmu}{\bm{\mu}}%
\newcommand{\bfnu}{\bm{\nu}}%
\newcommand{\bfxia}{\bm{\xi}}%
\newcommand{\bfpi}{\bm{\pi}}%
\newcommand{\bfrho}{\bm{\rho}}%
\newcommand{\bfsigma}{\bm{\sigma}}%
\newcommand{\bftau}{\bm{\tau}}%
\newcommand{\bfupsilon}{\bm{\upsilon}}%
\newcommand{\bfphi}{\bm{\phi}}%
\newcommand{\bfPhi}{\bm{\Phi}}%
\newcommand{\bfchi}{\boldsymbol{\chi}}%
\newcommand{\bfpsi}{\bm{\psi}}%
\newcommand{\bfomega}{\bm{\omega}}%
\newcommand{\bfvarepsilon}{\bm{\varepsilon}}%
\newcommand{\bfvartheta}{\bm{\vartheta}}%
\newcommand{\bfvarpi}{\bm{\vaarphi}}%
\newcommand{\bfvarrho}{\bm{\varrho}}%
\newcommand{\bfvarsigma}{\bm{\varsigma}}%
\newcommand{\bfvarkappa}{\bm{\varkappa}}%
\newcommand{\bfvarphi}{\bm{\varphi}}%
\newcommand{\Det}{\hbox{\rm det}\mskip2mu}
\newcommand{\cof}{\hbox{\rm cof}\mskip2mu}
\newcommand{\skw}{\hbox{\rm skw}\mskip2mu}
\newcommand{\sym}{\hbox{\rm sym}\mskip2mu}
\newcommand{\symz}{\text{sym}_0\mskip2mu}
\newcommand{\Tr}{\hbox{\rm tr}\mskip2mu}
\newcommand{\dev}{\text{dev}\mskip2mu}
\newcommand{\Rot}{\hbox{\rm Rot}\mskip2mu}
\newcommand{\Lin}{\hbox{\rm Lin}}
\newcommand{\Orth}{\hbox{\rm Orth}}
\newcommand{\rot}{\text{Orth}^+}
\newcommand{\trans}{\scriptscriptstyle\top}
\newcommand{\zed}{{\bf 0}}
\newcommand{\id}{{\bf 1}}
\newcommand{\Grad}{\hbox{\rm grad}\mskip2mu}
\newcommand{\Div}{\hbox{\rm Div}\mskip2mu}
\newcommand{\curl}{\hbox{\rm curl}\mskip2mu}
\newcommand{\Curl}{\hbox{\rm Curl}\mskip2mu}
\newcommand{\divx}{\text{div}\mskip2mu}
\newcommand{\grad}{\text{grad}\mskip2mu}
\newcommand{\B}{\text{B}}
\newcommand{\Bt}{\calB_t}
\newcommand{\dBt}{\partial\calB_t}
\newcommand{\p}{\text{P}}
\newcommand{\pt}{\calP_t}
\newcommand{\dpt}{\partial\calP_t}

\title{A Virtual Fields Method-Genetic Algorithm (VFM-GA) calibration framework for isotropic hyperelastic constitutive models with application to an elastomeric foam material}

\author{Zicheng Yan$^{1}$, Jialiang Tao$^{2}$, Christian Franck$^{2}$, and David L. Henann$^{1}$\footnotemark[2]\\
\\
\normalsize{{\it $^{1}$School of Engineering, Brown University, Providence, RI 02912, USA}}\\[0pt]
\normalsize{{\it $^{2}$Department of Mechanical Engineering, University of Wisconsin-Madison, Madison, WI 53706, USA}}}
\renewcommand*{\thefootnote}{\fnsymbol{footnote}}
\footnotetext[2]{Email address for correspondence: david\_henann@brown.edu}
\renewcommand*{\thefootnote}{\arabic{footnote}}
\date{}
\maketitle

\begin{abstract}
This work introduces a calibration framework for material parameter identification in isotropic hyperelastic constitutive models. The framework synergizes the Virtual Fields Method (VFM) to define an objective function with a Genetic Algorithm (GA) as the optimization method to facilitate automated calibration. The formulation of the objective function uses experimental displacement fields measured from Digital Image Correlation (DIC) synchronized with load cell data and can accommodate data from experiments involving homogeneous or inhomogeneous deformation fields. The framework places no restrictions on the target isotropic hyperelastic constitutive model, accommodating models with coupled dependencies on deformation invariants and specialized functional forms with a number of material parameters, and assesses material stability, eliminating sets of material parameters that potentially lead to non-physical behavior for the target hyperelastic constitutive model. To minimize the objective function, a GA is deployed as the optimization tool due to its ability to navigate the intricate landscape of material parameter space. The VFM-GA framework is evaluated by applying it to a hyperelastic constitutive model for compressible elastomeric foams. The evaluation process entails a number of tests that employ both homogeneous and inhomogeneous displacement fields collected from DIC experiments on open-cell foam specimens. The results outperform manual fitting, demonstrating the framework's robust and efficient capability to handle material parameter identification for complex hyperelastic constitutive models. 
\end{abstract}

\section{Introduction}\label{sec:intro}
Hyperelastic models describe the nonlinear, large-deformation elastic behavior of materials using a constitutive equation for the free-energy density. Capturing the nonlinearity of the elastic response of certain materials can require complex physically-informed functional forms for the free-energy density or phenomenological fitting functions with coupled dependencies on different deformation invariants, often leading to a large number of material parameters and posing a challenge to calibration \citep[e.g.,][]{ogden1972incomp,ogden1972comp,storakers86,attard2004hyperelastic,marckmann2006comparison,landauer2019experimental}. In particular, our work is motivated by hyperelastic modeling of elastomeric foams \citep{landauer2019experimental}, which consist of an elastomeric matrix with a gas-filled pore space and are widely used for impact mitigation. The mechanical behavior of elastomeric foams involves large, reversible deformation, substantial compressibility, and strong coupling between the volumetric and distortional responses, necessitating a sophisticated hyperelasticity model to capture the response. Addressing the robust and efficient calibration of such complex hyperelastic models is a goal of this work.

The task of material parameter identification involves using experimental data and a constitutive model’s structure to determine a suitable set of parameters, so that model predictions effectively capture experimental data. Often, this simply involves fitting a model to experimental stress-strain data \citep[e.g.,][]{gendy2000nonlinear}. However, the use of Digital Image Correlation (DIC) technology \citep[e.g.,][]{sutton1983determination,chu1985applications,sutton2009image} in experimental mechanics provides full-field displacement data for use in calibration alongside load-cell data. A noteworthy approach to material parameter identification using full-field displacement data is the Virtual Fields Method (VFM) developed by Pierron and coworkers \citep[e.g.,][]{grediac2006virtual,promma2009application,pierron2010identification,pierron2012virtual}. VFM utilizes the weak formulation of the balance of linear momentum with the DIC displacement fields and load-cell data to recast the task of determining best-fit material parameters as a minimization problem. Beyond VFM, other important calibration methods that use DIC displacement data have been developed \citep{avril2008overview}, such as finite element model updating \citep{chen2024finite} and the constitutive equation gap method \citep{geymonat2002identification}, and recent works leverage neural networks as optimizers for constitutive model calibration from full-field data, utilizing their adaptive learning capabilities to refine model parameters \citep[e.g.,][]{song2023identifying,hamel2023calibrating}.

The aforementioned works are generally intended to be applied to a specified constitutive model. The present work follows this paradigm, but we note the recent advancement of data-driven methods in constitutive modeling \citep{fuhg2024review}. These innovations include model-free approaches that bypass the formulation of a material model in favor of a material data set \citep[e.g.,][]{kirchdoerfer2016data,ibanez2017data} and machine-learning-based methods that aim to use some type of neural network \citep[e.g.,][]{ghaboussi1998autoprogressive,liang2008neural,linden2023neural} or Gaussian process regression \citep[e.g.,][]{frankel2020tensor,fuhg2022local,frankel2022machine} as a surrogate for a constitutive model. Moreover, recent work has pursued automated discovery of interpretable constitutive models \citep[e.g.,][]{flaschel2021unsupervised,wang2021inference} from a library of candidates that are either specified by the user or automatically generated \citep{kissas2024language}, and significant progress has been made for slightly compressible hyperelastic materials. However, the development of material model libraries for highly compressible hyperelastic materials that display strong coupling between the volumetric and distortional responses is not as mature, motivating our approach of developing a framework based on a specified constitutive model. That said, a critical inspiration for our work comes from the research of \citet{flaschel2021unsupervised} and their subsequent works \citep[e.g.,][]{joshi2022bayesian,flaschel2023automated_brain}. Besides their data-driven methodology, they proposed a physics-based objective function based on the balance of linear momentum that, as in VFM, may be used to assess the suitability of a parameter set for a given constitutive model based on experimental displacement field and reaction force data. In this work, we follow a similar VFM-type, mechanics-based approach for formulating an objective function, albeit for a selected constitutive model. 

The purpose of this paper is to introduce a novel framework for the identification of material parameters for a given isotropic, hyperelastic constitutive model based on experimental inputs. The approach integrates concepts from VFM to define an objective function with a Genetic Algorithm (GA) as the optimization tool. Building on the VFM foundations in \citet{grediac2006virtual} and the innovations of \citet{flaschel2021unsupervised}, the objective function is formulated based on the weak form of the quasi-static balance of linear momentum and is split into two parts: a balance of internal forces, accounting for equilibrium within the body, and a balance of external forces, accounting for equilibrium on boundaries on which displacements are prescribed. The objective function links DIC displacement data and synchronized load cell data with a specified hyperelastic constitutive model and is formulated in an implicit manner with respect to the material parameters, posing no restriction on the functional forms of the constitutive equations. The objective function is then optimized by a GA to obtain a calibrated material parameter set. GA is a metaheuristic evolutionary optimization method that mimics the process of natural selection \citep{holland1992genetic,mitchell1998introduction,sivanandam2008genetic} and has been deployed in numerous studies on material parameter calibration for constitutive models \citep[e.g.,][]{furukawa1997inelastic,chawla2009characterization,fernandez2018genetic}. However, most previous works using GA for material parameter calibration have focused on optimization of an error norm defined through the difference between experimental and model stress-strain curves, ignoring the lateral response of the material under homogeneous simple compression/tension or full-field displacement data from DIC under inhomogeneous deformation. An exception is the work of \citet{zhang2017verification,zhang2020vivo}, which combined VFM with differential evolution to identify mechanical properties of human optic nerve head tissues. The present work further pursues the integration of VFM and GA. Our specific GA strategy uses Boltzmann Selection \citep{goldberg1990note,maza1993analysis,mahfoud2000boltzmann}, Elitism \citep{bandyopadhyay1995pattern,bhandari1996genetic,ahn2003elitism}, and insights from Multipopulation Genetic Algorithms \citep{gorges1990explicit,cantu1999topologies} to maximize its computational efficiency and its performance in calibrating complex hyperelasticity models. In addition, the stability of a constitutive model for a certain set of parameters is assessed by testing the ellipticity of the quasi-static governing equations \citep{dacorogna2001} in common states of deformation to ensure the robustness of the fitted parameters. 

To facilitate broad application, we implement the VFM-GA framework in two functionalities, which respectively handle experimental inputs of (1) engineering stress-strain and lateral-axial strain curves from homogeneous simple compression/tension and (2) full-field displacement fields and synchronized load cell data from DIC experiments involving inhomogeneous deformation. A crucial aspect of this work is the assessment of the VFM-GA framework using real experimental data. To this end, we apply both functionalities to calibrating the hyperelastic constitutive model for elastomeric foams developed in \citet{landauer2019experimental} using experimental data for several densities of Poron XRD foams (Rogers Corp., Rogers, CT). To evaluate the first functionality, engineering stress-strain and lateral-axial strain data from homogeneous simple compression/tension experiments \citep{landauer2019experimental} is used, and the calibrated material parameter sets are subjected to several validation tests of their predictive capability. The second functionality is evaluated in applications involving both homogeneous and inhomogeneous deformation, using full-field displacement data generated using the SpatioTemporally Adaptive Quadtree mesh (STAQ) DIC algorithm of \citet{Yang2022} synchronized with load-cell data. The results from the evaluation tests demonstrate the VFM-GA framework's accuracy and efficiency compared to manual fitting.

The remainder of this paper is organized as follows. The hyperelastic constitutive model of \citet{landauer2019experimental} is first summarized in Section~\ref{sec:model}. Then, the two components of the VFM-GA framework are discussed, namely, the formulation of the objective function based on ideas from VFM in Section~\ref{sec:VFM} and the specifics of the GA-based optimization framework in Section~\ref{sec:GA}. The two functionalities are discussed in Section~\ref{sec:implementations}, and Section~\ref{sec:results} presents applications of the VFM-GA framework to real experimental data for elastomeric foams. Specifically, the first functionality, using data from homogeneous simple compression/tension experiments, is applied and subjected to validation tests in Section~\ref{sec:1st}, and the second functionality, which uses full-field displacement data from DIC measurements along with load-cell data, is demonstrated in Section~\ref{sec:2nd}. A computational noise study is conducted for the second functionality in Section~\ref{sec:noise}, and, in Section~\ref{sec:conc}, we close with some concluding remarks.

\section{Hyperelastic constitutive model}\label{sec:model}
Our motivation for the development of a VFM-GA framework is the complexity of certain hyperelastic constitutive models designed to broadly capture the mechanical response of elastomeric materials under varied deformation scenarios. Such constitutive models may involve coupled dependencies on deformation invariants and specialized phenomenological functions with a number of material parameters. These factors result in a challenging manual process for material parameter estimation, which is a roadblock to using such constitutive models. The current framework is designed to reduce impediments to deploying any hyperelastic constitutive model, nearly automating the calibration process, while possessing the flexibility to deal with real experimental data from different types of tests. To demonstrate our calibration framework on a complex hyperelastic constitutive model, we apply it to the hyperelastic constitutive model for compressible elastomeric foams, developed in our prior work \citep{landauer2019experimental}. The details of this constitutive model are summarized in this section prior to describing the calibration framework in the subsequent sections.

\paragraph{Basic kinematics}  We begin by denoting the region of space occupied by a body in the fixed reference configuration as $\mathcal{B}$. Within $\mathcal{B}$, $\mathbf{X}$ denotes an arbitrary material point. The reference body $\mathcal{B}$ then undergoes a motion $\bfx = \bfchi(\mathbf{X}, t)$ to the deformed body $\Bt$ at each time $t$, where $\mathbf{x}$ denotes the spatial point within $\Bt$ corresponding to the material point $\bfX$. The deformation gradient is $\bfF = \nabla\bfchi$ with the ratio between the deformed and reference volumes strictly greater than zero, i.e., $J = \Det\bfF > 0$.\footnote{{\it Notation:} The symbols $\nabla$, Div and Curl denote the gradient, divergence, and curl with respect to the material point $\mathbf{X}$ in the reference body $\mathcal{B}$; grad, div, and curl denote these operators with respect to the point $\bfx=\bfchi(\mathbf{X},t)$ in the deformed body $\Bt$. We write $\Tr\bfA$, $\det\bfA$, $\dev\bfA$, $\sym\bfA$, and $\skw\bfA$ to denote the trace, determinant, deviatoric part, symmetric part, and skew part of a tensor $\bfA$, respectively.}  The displacement field is $\bfu(\mathbf{X},t) = \bfchi(\mathbf{X},t)-\mathbf{X}$. We make use of the left polar decomposition of the deformation gradient, $\bfF = \bfV\bfR$, where $\bfV$ is the left stretch tensor and $\bfR$ is the rotation tensor. The spectral decomposition of $\bfV$ is $\bfV = \sum_{i=1}^3(\lambda_i)\bfl_i\otimes\bfl_i$, where $\{\lambda_i|i=1,2,3\}$ are the principal stretches and $\{\bfl_i|i=1,2,3\}$ are the principal directions of $\bfV$. The spatial logarithmic (Hencky) finite-strain tensor is given by $\bfE = \sum_{i=1}^3(\ln\lambda_i)\bfl_i\otimes\bfl_i$.

\paragraph{Free-energy function} For an isotropic, hyperelastic material, the referential free-energy density $\psi$ may be given as a function of three invariants $\{K_1, K_2, K_3\}$ of the spatial logarithmic strain tensor $\bfE$, as proposed by \citet{criscione2000}. The first invariant $K_1 =  \Tr(\bfE) = \ln(J)$ describes pure volumetric change, and the second invariant $K_2 = |\dev(\bfE)| \ge 0$ corresponds to the magnitude of constant-volume distortion. The third invariant $K_3 = 3\sqrt{6}\,\Det(\bfN) \in [-1, 1]$ represents the mode of constant-volume distortion, where $\bfN = \dev(\bfE)/K_2$ is the tensorial direction of deviatoric strain. The invariant $K_3$ is constant for a given mode of deformation, e.g., $K_3 = -1$ for simple compression, $K_3 = 0$ for pure shear, and $K_3 = 1$ for simple tension. As shown in \citet{criscione2000}, the Cauchy stress $\bfT$ may be obtained through the derivatives of the free-energy density function $\psi = \tilde\psi(K_1,K_2,K_3)$ as follows: 
\begin{equation}\label{eq:cauchy}
\bfT = J^{-1}\left[\dfrac{\partial \tilde\psi}{\partial K_1}\id + \dfrac{\partial\tilde\psi}{\partial K_2}\bfN +\dfrac{\partial\tilde\psi}{\partial K_3}\dfrac{1}{K_2}\bfY\right],
\end{equation}
where $\bfY = 3\sqrt{6}\bfN^2 - \sqrt{6}\id - 3K_3\bfN$.

\paragraph{Decomposition of the free-energy density function} We take the free-energy density function $\tilde\psi(K_1,K_2,K_3)$ to be additively decomposed as follows: 
\begin{equation}\label{eq:free_energy}
\tilde\psi(K_1,K_2,K_3) = G_0 \left[ X(K_1) K_2^2 + L(K_2, K_3) \right] + B f(K_1),
\end{equation}
where $G_0$ and $B$ are the ground-state, equilibrium shear and bulk moduli, respectively, and $X(K_1)$, $L(K_2,K_3)$, and $f(K_1)$ are phenomenological fitting functions. These phenomenological functions are crucial to describe the nonlinear response of elastomeric foams across compression, tension, and shear. The $X(K_1) K_2^2$ term characterizes the coupling between the volumetric and distortional responses, such that the function $X(K_1)$ may be interpreted as describing the volumetric-strain-dependence of the instantaneous shear modulus. The $G_0 L(K_2, K_3)$ term is responsible for the higher-order distortional response and is taken to be uncoupled from volume change, and the $B f(K_1)$ term captures the purely volumetric response. Using \eqref{eq:free_energy} in \eqref{eq:cauchy}, we obtain the following expression for the Cauchy stress:
\begin{equation}\label{eq:cauchy2}
\bfT = J^{-1}\left[\left(G_0\dfrac{dX}{dK_1}K_2^2 + B\dfrac{d f}{dK_1}\right){\bf 1} + G_0\left(2X(K_1)K_2 + \dfrac{\partial L}{\partial K_2}\right)\bfN + G_0\dfrac{\partial L}{\partial K_3}\dfrac{1}{K_2}\bfY\right].
\end{equation}
Specific forms for the phenomenological fitting functions $X(K_1)$, $L(K_2,K_3)$, and $f(K_1)$ may then be adopted for different foam materials. 

\paragraph{Specialization of the free-energy function}\label{para:spec}
The three phenomenological functions in the free-energy density function $\tilde\psi$ \eqref{eq:free_energy} adopted here are suitable for a wide range of elastomeric foams, both open-cell and closed-cell, with a typical compression response that involves linear, plateau, and densification regimes. Specific forms are given below, and further details are discussed in \citet{landauer2019experimental}.
\begin{enumerate}
\item The distortional/volumetric-coupling response function $X$:
\begin{equation}\label{eq:coupled_fit}
X( K_1 ) = \dfrac{1}{2} \left( X_1^\prime + X_2^\prime \right) K_1 +  \dfrac{\Delta_{\rm K}}{2} \left( X_1^\prime - X_2^\prime \right) \ln \left[\dfrac{ \cosh \left( (K_1 - K_1^0)/\Delta_{\rm K} \right) }{\cosh \left(K_1^0/\Delta_{\rm K} \right)}\right] + 1.
\end{equation}
The function $X(K_1)$ involves a linear dependence on $K_1$ with different slopes before and after the transition from the linear regime to the plateau regime in a compression load path. The parameters $X_1^\prime$ and $X_2^\prime$ represent the limiting values of $dX/dK_1$ prior to and after the transition to the plateau regime, respectively. The parameter $K_1^0 < 0$ denotes the value of $K_1$ at which the transition occurs, and the parameter $\Delta_{\rm K} > 0 $ corresponds to the range of $K_1$ over which the transition takes place.

\item The higher-order distortional response function $L$:
\begin{equation}\label{eq:L_1}
L (K_2,K_3) = C_0 K_{2}^{p} + C_1 \left( 1 + K_3 \right) K_2^{q},
\end{equation}
where $\{C_0 > 0,p>2\}$ and $\{C_1>0,q>2\}$ are two pairs of constant parameters. The function $L(K_2,K_3)$ is responsible for capturing the higher-order, large-deformation distortional response, uncoupled from the volumetric response. Note that in simple compression, $K_3 = -1$, and the second term vanishes, allowing different higher-order distortional responses in tension and compression to be captured. 

\item The purely volumetric response function $f$:
\begin{equation}\label{eq:f_vr}
f(K_1) = \dfrac{J^{C_2} - C_2\ln J -1}{C_2^2} + \dfrac{C_3}{r-1}\left[  \dfrac{-1}{J^{r-1}} + \dfrac{(1-J_{\rm min})^r}{(J-J_{\rm min})^{r-1}} + J_{\rm min}\right],
\end{equation}
where $J = \exp(K_1)$. The function $f(K_1)$ is dedicated to describing the purely volumetric response. The first term, involving the constant parameter $C_2$, is introduced to capture the volumetric response during the transition between the linear regime and the plateau regime in compression. The term involving the constant parameter $C_3$ is introduced to capture the rapidly stiffening behavior in the large-strain densification regime in compression. The constant parameter $0<J_{\rm min}<1$ represents the minimum limiting value of $J$ at which the divergence of $df/dK_1$ occurs as $J$ approaches $J_{\rm min}$ from above, and the constant parameter $r$ controls the rate of this divergence, enabling the shape of the stress-strain curve to be captured in the densification regime. 
\end{enumerate}
In summary, the constitutive model described above involves strong coupling between the first and second invariants of the logarithmic strain, specialized phenomenological fitting functions that are mathematically complex, and a set of 14 material parameters $\boldsymbol{\theta} = \{ G_{0}$, $B$, $J_{\rm min}$, $C_1$, $K_1^{0}$, $\Delta_K$, $X_1^\prime$, $X_2^\prime$, $C_0$, $p$, $q$, $C_2$, $C_3$, $r\}$---the influence of which cannot be additively decoupled. The challenges faced in manual calibration of this model make it a suitable test case for the VFM-GA framework described in the subsequent sections. 

\section{Mechanics Framework: Virtual Fields Method}\label{sec:VFM}
\subsection{Overview}\label{sec:mechanics1}
\begin{figure}[!t]
    \centering
    \includegraphics[width=1\textwidth]{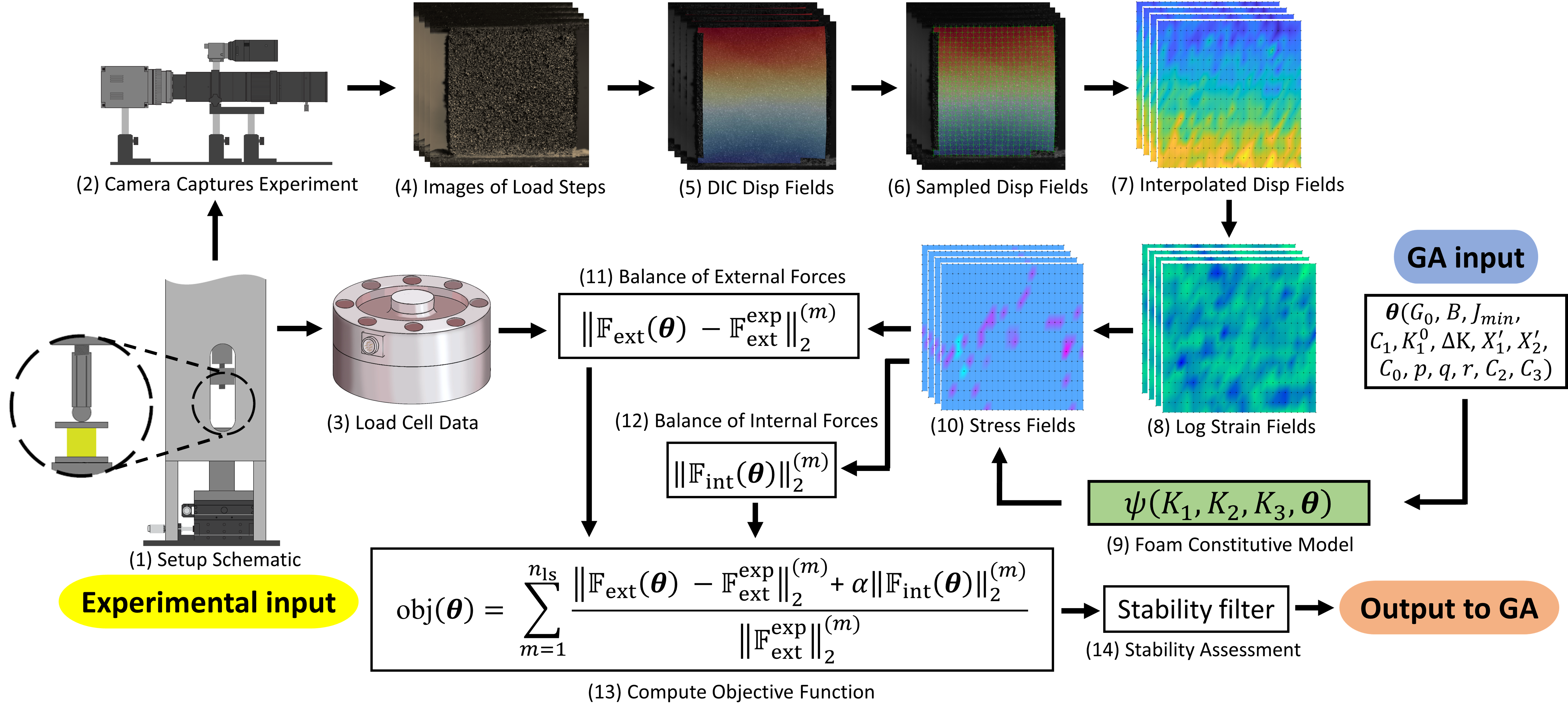}
    \caption{Schematic of the mechanics framework based on the Virtual Fields Method (VFM). The inputs for the mechanics framework are a material parameter set $\boldsymbol{\theta}$ and experimental DIC and load cell data, and the outputs are an objective function value, which may be used in optimization. Starting from the experimental input side (left), (1) shows a typical experimental setup. In (2), a camera images the specimen over the experimental load path, and, in (3), a load cell collects reaction force data synchronized with the camera. The camera outputs images at each load step with index $m$ in (4). With a DIC algorithm, displacement fields may be calculated in (5) and sampled using an undeformed mesh in (6). The sampled displacement fields are interpolated using shape functions in (7), and deformation gradient fields are computed in (8) to obtain logarithmic strain fields. Then, starting from the right side, we have (9) where a given material parameter set $\boldsymbol{\theta}$ is plugged into the designated hyperelastic constitutive model $\psi$. Combining (8) and (9), stress fields at each load step $m$ are then computed in (10). Combining load cell data (3) and the computed stress fields (10), the balances of (11) external forces and (12) internal forces may be computed at each load step $m$. In (13), the two balance residuals are combined, normalized, and summed over the load step index $m$ to compute the objective function ${\rm obj}(\boldsymbol{\theta})$. Before passing the objective function value into the optimization method (here, a genetic algorithm), it passes through a stability filter (14) where material stability associated with the given material parameter set is assessed.}
    \label{fig:framework}
\end{figure}

The goal of the VFM-GA framework is to find a suitable material parameter set for a given constitutive model to capture a given experimental dataset. In this section, we present the mechanics framework that is used to compute an objective function, which may then be minimized by a compatible optimization method. The mechanics framework utilizes concepts from the Virtual Fields Method \citep{grediac2006virtual,promma2009application,pierron2010identification,pierron2012virtual} and is formulated based on the quasi-static balance of linear momentum. Inspired by the formulation of the objective function in \citet{flaschel2021unsupervised}, the contribution of the quasi-static balance of linear momentum to the objective function is split into two parts: a balance of internal forces, accounting for equilibrium within the body, and a balance of external forces, accounting for equilibrium on boundaries on which displacements are prescribed. An overview schematic of the workflow for calculating the objective function is illustrated in Fig.~\ref{fig:framework}, and each step is discussed in detail in the subsequent sections. While the schematic illustrates the method applied to our constitutive model for elastomeric foams, the method does not place restrictions on the specific form of the hyperelastic constitutive model.

\subsection{DIC kinematics and data processing}\label{kinematics}
Illustrated in Fig.~\ref{fig:framework}, parts (1)-(6), the mechanics framework begins with inputs from experimental data. On the left-hand side of Fig.~\ref{fig:framework}, we have an experimental setup (part (1)), illustrated for but not restricted to compression, in which a camera records images of one face of the specimen (part (2)) and synchronized reaction forces are measured by a load cell (part (3)). The camera records images of the specimen in the reference configuration and in the deformed configuration at load steps $m\in\{1,2,\ldots,n_{\rm ls}\}$ (part (4)), where $n_{\rm ls}$ is the total number of load steps, which are then passed to a DIC algorithm. In this work, we focus on experiments in which the displacement field may be idealized as two-dimensional in the $X_1$-$X_2$-plane. A specific DIC algorithm is utilized to obtain the full-field, two-dimensional displacement fields at each load step $m$ throughout the load path of the experiment (part (5)). In this study, we use the STAQ-DIC method of \citet{Yang2022}. We note that other DIC algorithms, possessing similar functionalities, may be adopted. The guidelines for choosing a DIC algorithm compatible with the mechanics framework are that (1) the region of interest (ROI) must deform with the specimen and (2) it must be capable of calculating displacements sufficiently close to surfaces. After the DIC algorithm processes the images and obtains displacement fields for each load step $m$, the undeformed ROI is spatially discretized and a mesh is created, involving nodes $A\in\{1,2,\ldots,n_{\rm n}\}$, where $n_{\rm n}$ is the total number of nodes. The DIC displacement fields for each load step $m$ are then sampled at the reference coordinates $(X_1^A,X_2^A)$ corresponding to each node $A$, resulting in a two-dimensional displacement $\bfu^A$ at each node $A$ for each load step $m$ (part (6)). 

\subsection{Computation of the deformation gradient} 
Next, the two-dimensional components of the displacement field $\bfu(X_1,X_2)$ for each load step $m$ are interpolated from the nodal displacements $\bfu^{\rm A}$ using the mesh as follows (part (7)):
\begin{equation}\label{disp_eqn}
u_i(X_1,X_2) = \sum_{A=1}^{n_{\rm n}} u_i^A N^A(X_1,X_2),\quad i\in\{1,2\},
\end{equation}
where $N^A(X_1,X_2)$ are the global shape functions for each node $A$. The two-dimensional components of the deformation gradient $\bfF$ follow as  
\begin{equation}\label{disp_eqn2}
    F_{ij}(X_1,X_2) = \delta_{ij} + \sum_{A=1}^{n_{\rm n}}  u^A_i\dfrac{\partial N^A}{\partial X_j},\quad i,j\in\{1,2\},
\end{equation}
where $\delta_{ij}$ is the Kronecker delta. Consistent with our focus on two-dimensional problems, we assume that $F_{13}=F_{31}=F_{23}=F_{32}=0$ but allow $F_{33}(X_1,X_2)\ne1$, so that both plane-strain and plane-stress approximations may be accommodated. Under this idealization, the two-dimensional components of the spatial logarithmic finite-strain tensor $\bfE$ may be calculated (part (8)).

\subsection{Computation of stress fields}\label{sec:stress_field}
Given a hyperelastic constitutive model (part (9)), the next step is to calculate the two-dimensional components of the Cauchy stress field (part (10)). To do so, it is necessary to determine the out-of-plane component of the deformation gradient $F_{33}$. Here, we consider two possible treatments: plane-strain and plane-stress approximations. In a plane-strain approximation, one simply takes $F_{33}=1$ to be spatially uniform. This is the most straightforward approach to implement in the mechanics framework but is often not valid in experiments, requiring either that a prismatic specimen be constrained between parallel, flat, frictionless walls along the $X_3$-direction or that a prismatic specimen have an out-of-plane thickness that is much greater than its in-plane dimensions. Either approach presents a challenge to obtaining images of two-dimensional displacement fields on the end face of the specimen. For the first approach, even if lubricated, clear walls are used, friction between the specimen and the wall can affect the displacement field on the end face. For the second approach, even for a sufficiently thick specimen, attaining a spatially uniform value of $F_{33}=1$ is challenging. For example, in our own tests using a specimen with a square cross-section and an out-of-plane depth eight times the in-plane dimensions, out-of-plane barreling of the end face in compression led to inaccurate two-dimensional displacement fields calculated by the DIC algorithm. Accordingly, the plane-strain approximation is not employed in the applications discussed in this paper. 

Alternatively, in a plane-stress approximation, $F_{33}$ at each point may be determined from the requirement that the out-of-plane normal Cauchy stress component is zero, i.e., $T_{33}=0$. For the constitutive model of Section~\ref{sec:model}, the components of the Cauchy stress tensor \eqref{eq:cauchy2} are calculated using the in-plane components of $\bfF$ and the unknown value of $F_{33}$. The requirement that $T_{33}=0$ yields a nonlinear equation for $F_{33}$, which we then solve using a mixed bisection/Newton-Raphson approach. Once $F_{33}$ has been obtained, the in-plane stress components may be calculated. The calculation of $F_{33}$ under the plane-stress approximation is performed at integration points and is discussed further in Section~\ref{fem_details}. Based on our experience in numerical tests, optimized material parameter sets may be more accurately obtained under the plane-stress approximation than the plane-strain approximation, even for specimens with aspect ratios of order one. Therefore, the plane-stress approximation is used throughout the illustrative examples presented in this paper.

\subsection{Computation of the objective function}  
The workflow for computing the objective function is illustrated in Fig.~\ref{fig:framework}, parts (11)-(13). The objective function is based on the quasi-static balance of linear momentum, which, following \citet{flaschel2021unsupervised}, is split into two contributions, which we refer to as the balances of internal and external forces. Denoting the first Piola-Kirchhoff stress as $\bfP = J\bfT\bfF^{-\top}$, equilibrium in the reference body $\mathcal{B}$ requires that ${\rm Div}\,\bfP + \bfb_{\rm R} = {\bf 0}$, where $\bfb_{\rm R}$ is the body force per unit reference volume. Let $\mathcal{S}_1$ and $\mathcal{S}_2$ denote complementary subsurfaces of the boundary $\partial\mathcal{B}$ ($\mathcal{S}_1 \cup \mathcal{S}_2 = \partial\mathcal{B}$, $ \mathcal{S}_1 \cap \mathcal{S}_2 = \varnothing$), on which displacement and traction boundary conditions are applied, respectively. The boundary conditions on $\partial\mathcal{B}$ are $\bfu = \hat\bfu$ on $\mathcal{S}_1$ and $\bfP\bfn_{\rm R} = \hat\bft_{\rm R}$ on $\mathcal{S}_2$, where $\bfn_{\rm R}$ is the outward unit normal vector on $\mathcal{S}_2$, and $\hat\bfu$ and $\hat\bft_{\rm R}$ are prescribed. 

We introduce a vector test function $\bfw(X_1,X_2)$, i.e., the virtual field, from the set $\mathbb{V} = \{\bfw|\bfw \in [H^1(\mathcal{B})]^2,\bfw={\bf 0}\text{ on }\mathcal{S}_1\}$, where $H^1$ denotes the Sobolev space of degree one, and write the weak form of the equilibrium equations as 
\begin{equation}\label{principle of virtual work 2}
\int_{\mathcal{B}} \bfP:\dfrac{\partial{\bfw}}{\partial{\bfX}} \,dv = \int_{\mathcal{B}} {\bf{b}_{\rm R} \cdot \bf{w}} \,dv + \int_{\mathcal{S}_2} \bft_{\rm R}  \cdot \bfw \,da,\quad \forall \bfw\in\mathbb{V}.
\end{equation}
Henceforth, we neglect body forces, so that $\bfb_{\rm R}=\zed$ in $\mathcal{B}$, and only consider problems in which displacement-controlled loading is prescribed, so that ${\bf t}_{\rm R}=\zed$ on $\mathcal{S}_2$, and the right-hand-side of \eqref{principle of virtual work 2} vanishes.

Next, following the approach of \citet{flaschel2021unsupervised}, let $\mathcal{D} = \{(A,i)|A=1,2,\ldots,n_{\rm n};i=1,2\}$ be the set of all global degrees of freedom, and denote the subset of degrees of freedom on which the displacement $\hat{u}_i^A$ is prescribed as $\mathcal{D}_{\rm ext}\subseteq \mathcal{D}$. The subset of degrees of freedom on which displacements are not prescribed is then $\mathcal{D}_{\rm int}=\mathcal{D}\setminus\mathcal{D}_{\rm ext}$. As in a Galerkin approach, we consider a subset of piecewise test functions generated through the mesh, in which the test function is interpolated using the same shape functions as the displacement field:
\begin{equation}\label{delta v}
w_i(X_1,X_2) = \sum_{A=1}^{n_{\rm n}}w_i^A N^A(X_1,X_2),\quad i\in\{1,2\},
\end{equation}
with $w_i^A=0$ if $(A,i)\in\mathcal{D}_{\rm ext}$. Using \eqref{delta v} in the weak form of equilibrium \eqref{principle of virtual work 2}, we are led to 
\begin{equation}\label{balance of internal force}
\int_{\mathcal{B}} P_{ij}\frac{\partial N^A}{\partial X_j} \,dv  = 0, \quad \forall  (A,i)\in \mathcal{D}_{\rm int},
\end{equation}
since $w_i^A$ for $(A,i)\in\mathcal{D}_{\rm int}$ are arbitrary. Since the mesh and the displacement \eqref{disp_eqn} and deformation gradient \eqref{disp_eqn2} fields are fixed for a given load step $m$, \eqref{balance of internal force} is a function of only the material parameter set $\boldsymbol{\theta}$ of the target constitutive model through the first Piola-Kirchhoff stress $\bfP$. After numerical evaluation of the integrals in \eqref{balance of internal force}, we obtain the balance of internal forces:
\begin{equation}\label{balance of internal force 2}
\mathbb{F}_{\mathrm{int}}(\boldsymbol{\theta}) = 0,
\end{equation}
where $\mathbb{F}_{\mathrm{int}}$ is a vector of length $|\mathcal{D}_{\rm int}|$, in which each component represents a residual force for the corresponding global degree of freedom in $\mathcal{D}_{\rm int}$. 

As discussed in \citet{flaschel2021unsupervised}, the nodal reaction forces corresponding to each prescribed degree of freedom in $\mathcal{D}_{\rm ext}$ are not experimentally accessible. Instead, only the sums of the nodal reaction forces for complementary subsets of $\mathcal{D}_{\rm ext}$, each corresponding to a set of nodes on a subsurface of $\mathcal{S}_1$ and one coordinate direction, are experimentally available from load-cell data. With $n_{\rm s}$ the total number of subsets, denote each subset of $\mathcal{D}_{\rm ext}$ as $\mathcal{D}_{\rm ext}^k$ with $k\in\{1,2,\ldots,n_{\rm s}\}$, $\cup_{k=1}^{n_{\rm s}}\mathcal{D}_{\rm ext}^k = \mathcal{D}_{\rm ext}$, and $\mathcal{D}_{\rm ext}^k\cap\mathcal{D}_{\rm ext}^l=\varnothing$ for $l\ne k$. We note that each subset $k$ only involves one coordinate direction, i.e., $i$ is the same for all $(A,i)\in\mathcal{D}^k_{\rm ext}$. Then, the total reaction force for subset $k$ is
\begin{equation}\label{balance of external force 2}
\sum_{(A,i) \in \mathcal{D}_{\rm ext}^{k}} \int_{\mathcal{B}} P_{ij}  \dfrac{\partial N^A}{\partial X_j}  \,dv = R^{k}, \quad \forall k\in\{1,2,\ldots,n_{\rm s}\},
\end{equation}
where $R^k$ is known from experimental load-cell data. As for the balance of internal forces in the preceding paragraph, \eqref{balance of external force 2} is a function of only the material parameter set $\boldsymbol{\theta}$ through the stress $\bfP$. After numerical evaluation of the integrals in \eqref{balance of external force 2}, we obtain the balance of external forces:
\begin{equation}\label{balance of external force 3}
\mathbb{F}_{\mathrm{ext}}(\boldsymbol{\theta}) = \mathbb{F}_{\mathrm{ext}}^{\mathrm{exp}},
\end{equation}
where $\mathbb{F}_{\mathrm{ext}}$ and $\mathbb{F}_{\mathrm{ext}}^{\mathrm{exp}}$ are vectors of length $n_{\rm s}$. The components of $\mathbb{F}_{\mathrm{ext}}$ represent the computed net reaction force for each subset $k$ as a function of $\boldsymbol{\theta}$, and the components of $\mathbb{F}_{\mathrm{ext}}^{\mathrm{exp}}$ represent the corresponding reaction force components experimentally measured by a load cell. 

Together, \eqref{balance of internal force 2} and \eqref{balance of external force 3} comprise a nonlinear and overdetermined system of equations for the material parameter set $\boldsymbol{\theta}$. Here, we use \eqref{balance of internal force 2} and \eqref{balance of external force 3} to construct an objective function that may be minimized with respect to $\boldsymbol{\theta}$ to yield an optimal material parameter set. Considering DIC displacement data and corresponding load-cell reaction force data at $n_{\rm ls}$ load steps, the $L^2$ norms of the balances of internal and external forces for each load step $m$ are combined, normalized, and summed up over all load steps to obtain the following objective function: 
\begin{equation}\label{obj}
{\rm obj}(\boldsymbol{\theta}) = \sum_{m=1}^{n_{\rm ls}} \dfrac{\|{\mathbb{F}_{\mathrm{ext}}(\boldsymbol{\theta})-\mathbb{F}_{\mathrm{ext}}^{\mathrm{exp}}\|^{(m)}_2 + \alpha \|{\mathbb{F}_{\mathrm{int}}(\boldsymbol{\theta})}\|_2^{(m)}}}{\|\mathbb{F}_{\mathrm{ext}}^{\mathrm{exp}}\|^{(m)}_2 }.
\end{equation}
The contribution of the internal force balance relative to the external force balance for a single load step is weighted by the parameter $\alpha$, the effect of which is revisited in Section~\ref{sec:noise}. The contributions to the objective function due to each load step $m$ are normalized by the respective norm of the experimentally measured reaction forces $\|\mathbb{F}_{\mathrm{ext}}^{\mathrm{exp}} \|_2^{(m)}$. Without such a normalization, the contribution to the objective function for load steps corresponding to comparably large deformations and high stresses would tend to be much greater than those for load steps corresponding to comparably small deformations and low stresses. Thus, the normalization factor is necessary to ensure that each load step has approximately equal weight in the objective function and to obtain an optimized material parameter set that is capable of capturing the material response throughout the experimental load path. 

\subsection{Comments on the mesh and numerical integration}\label{fem_details}
In our implementation of the VFM framework, we utilize two-dimensional, four-node, quadrilateral elements to construct the undeformed mesh since this element type is readily compatible with the DIC algorithm used in this study. To numerically evaluate the integrals \eqref{balance of internal force} and \eqref{balance of external force 2}, we employ full integration with four integration points in most scenarios. Additionally, several other types of elements and numerical integration procedures have been implemented and tested, including a reduced quadrature method for nonlinear solids \citep{li2022reduced,li2023improved}, which helps reduce computational cost, while resulting in errors that are consistently below 1\% compared to full integration during objective function computation. 

To compute the stress at each integration point under the plane-stress approximation, it is necessary to solve a nonlinear equation, stemming from the requirement that $T_{33}=0$, for the out-of-plane deformation gradient component $F_{33}$. Our approach is to use the bisection method to solve for $F_{33}$ for the first integration point in a given load step. Then, since integration points within the same spatial vicinity are expected to have similar values of $F_{33}$, the Newton-Raphson method is used to solve for $F_{33}$ for all subsequent integration points in the same load step, in which the initial guess for $F_{33}$ is the solution from the prior integration point. In our experience, this approach combines the guaranteed convergence of the bisection method with the efficiency of the Newton-Raphson method to reduce computational cost while preserving robustness. 

\subsection{Stability filters}\label{sec_stability}
After formulating the objective function, one can move directly to optimization if the constitutive model under consideration yields stable behavior over the specified ranges of the material parameters. However, as our VFM-GA framework is specifically designed to tackle hyperelastic constitutive models with couplings between deformation invariants and complex phenomenological functions, certain combinations of material parameters can lead to unstable behavior. For instance, for the compressible hyperelastic model of Section \ref{sec:model}, instability can occur in the plateau regime in compression. To avoid this issue, stability filters, indicated by Fig.~\ref{fig:framework}, part (14), are implemented in the mechanics framework to identify undesirable material parameter sets and improve the robustness of the optimization procedure. These filters are implemented to be consistent with a GA optimization method of Section~\ref{sec:GA}. 

\subsubsection{Loss of ellipticity filter}\label{sec:loss_of_ellipticity}
The first filter is based on the ellipticity of the quasi-static governing equations, in which loss of ellipticity is identified as the condition at which localization of deformation becomes possible \citep[e.g.,][]{triantafyllidis1986gradient,bigoni2012}. For a given material parameter set, we test for the loss of ellipticity at several homogeneous base states of deformation using the necessary and sufficient conditions of \citet{dacorogna2001}. For an isotropic, hyperelastic material, the homogeneous base state may be identified by a set of invariants $\{K_1,K_2,K_3\}$, and, to balance efficiency and robustness, the base states for checking ellipticity should be judiciously chosen for the constitutive model under consideration. The model of Section~\ref{sec:model} is vulnerable to instability during the plateau regime in compression \citep{landauer2019experimental}, so we consider five check points with $K_3=-1$ and $(K_1,K_2)=(-0.15,0.15),(-0.25,0.25),(-0.35,0.35),(-0.45,0.45)$, and $(-0.55,0.55)$. Additionally, we also consider one check point in shear with $K_3=0$ and $(K_1,K_2)=(0,0.51)$ to ensure that material behavior is stable to large deformation in shear. Then, for each check point, the material moduli $\partial^2\psi/\partial F_{ij}\partial F_{kl}$ are calculated, and the necessary and sufficient conditions for ellipticity of \citet{dacorogna2001} are tested. If loss of ellipticity is detected, the computed objective function for that parameter set is overwritten by a large hard-coded value, effectively filtering it out in the GA-based optimization. Since we do not exhaustively test across all base states of deformation, it is possible that a material parameter set that passes the ellipticity filter can still exhibit unstable behavior under other base states of deformation. That said, we find that the choice of check points employed here effectively identifies material parameter sets that lead to unstable behavior for the constitutive model of Section~\ref{sec:model}. For other constitutive models, check points should be carefully chosen based on the characteristics of the specific model. 

\subsubsection{Monotonicity filters for simple compression/tension}\label{sec:monotonic}
The loss of ellipticity filter only depends on the given material parameter set and the choice of check points, so, in our framework, it is applied for any type of input experimental data. In contrast, we also implement filters that are only applied when the input data is from simple compression/tension experiments, which check the monotonicity of the axial engineering stress and volumetric strain responses over the simple compression/tension load path. The monotonicity filters are implemented by first sorting the input load steps in terms of ascending axial engineering strain magnitude for simple compression and tension, respectively. Then, for a given material parameter set, the calculated axial engineering stress magnitude and volumetric strain magnitude $|K_1|$ for a given load step are compared to the respective quantities for the previous load step with a lower axial engineering strain magnitude. If the magnitude of axial engineering stress or volumetric strain is found to be lower in the load step, non-monotonic behavior is detected, and that material parameter set is filtered out by overwriting the computed objective function by a large hard-coded value. 

\section{Optimization framework: Genetic algorithm}\label{sec:GA}
Following the discussion of the VFM part of the VFM-GA framework, we now shift our focus to the GA part of the framework, namely, applying a genetic algorithm as the optimization tool to minimize the objective function ${\rm obj}(\boldsymbol{\theta})$ with respect to the material parameter set $\boldsymbol{\theta}$. 

\subsection{The choice of optimization method}
Before diving into the details of the GA, we articulate the rationale behind the choice of optimization method. Regarding gradient-based optimization methods, these methods typically start with an initial guess, approaching optimized results using the computed gradients of the objective function with respect to the optimization variables. Functionally, this category of methods is compatible with our objective function in the VFM framework (except the stability filters of Section~\ref{sec_stability}), but, in our empirical experience with gradient-based methods, we observed that in a highly dimensioned material parameter space (e.g., 14 parameters in the model of Section \ref{sec:model}), the landscape of the objective function is intricate. There are other limiting factors to consider---e.g., the challenge of identifying an appropriate initial guess and the cost of computing tangents---which escalate with the number of material parameters. These limitations compel us to look beyond gradient-based methods.

Ultimately, we chose an optimization method from a class of evolutionary algorithms that are metaheuristic, recognizing their robust capacity to navigate complex optimization landscapes. Examples of evolutionary algorithms are Genetic Algorithms (GA) \citep[e.g.,][]{holland1992genetic,mitchell1998introduction,sivanandam2008genetic}, Particle Swarm Optimization (PSO) \citep[e.g.,][]{kennedy1995particle,poli2007particle}, and Differential Evolution (DE) \citep[e.g.,][]{das2010differential,price2013differential}. We adopt GA as the optimization method in our VFM-framework due to its compatibility with the objective function, its global search capability, and its amenability to parallelism in computation. In general, genetic algorithms mimic the process of natural selection in optimization, and act on optimization variables within a prescribed range. While the other two evolutionary algorithms mentioned above are also viable in principle, we did not investigate these alternatives in this study. The schematic of the GA implemented in our work is shown in Fig.~\ref{fig:ga_schematic}. The details of the implementation are explained in the following subsections. 

\subsection{Genetic algorithm implementation}
\begin{figure}[!t]
    \centering
    \includegraphics[width=1\textwidth]{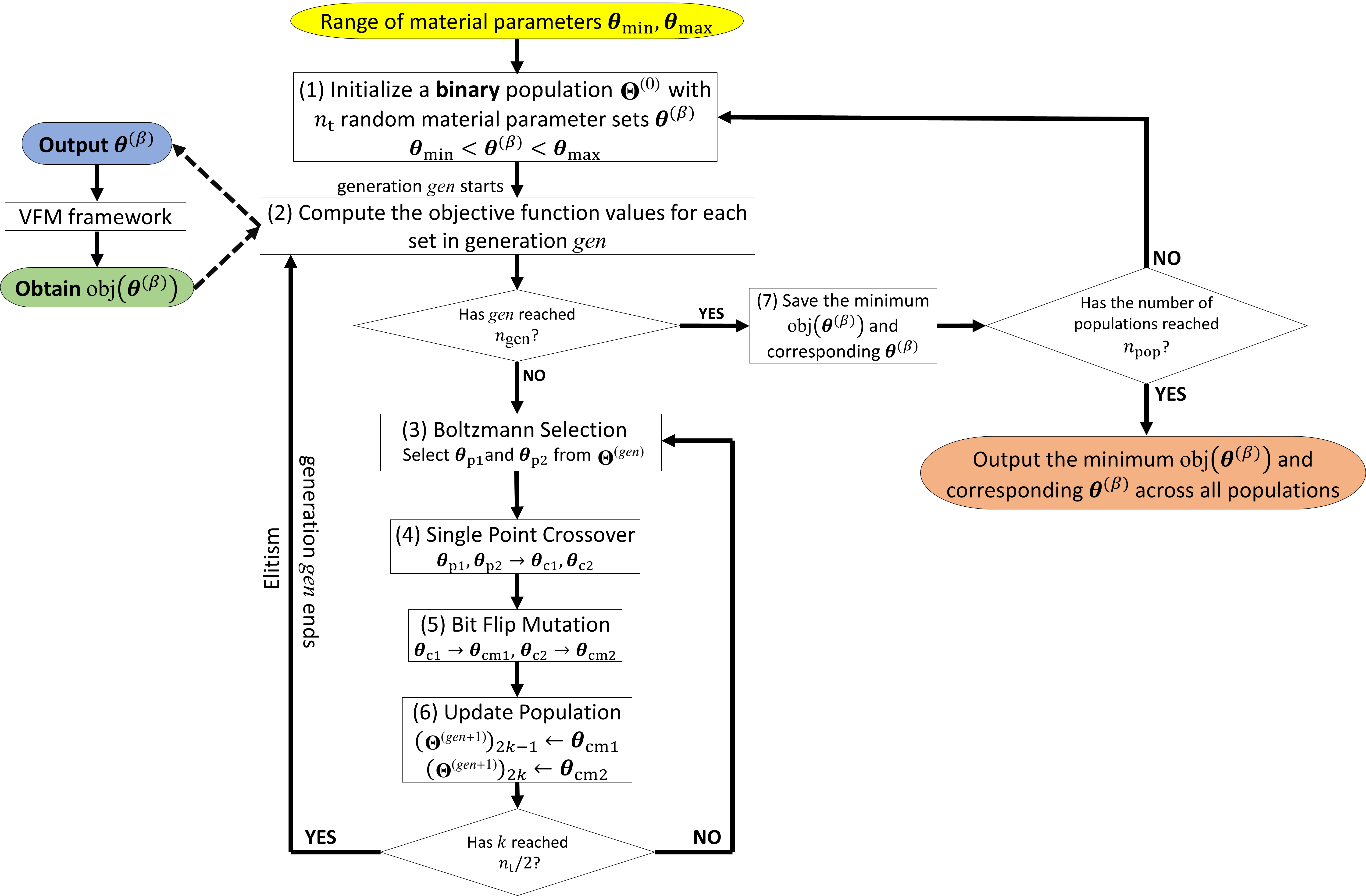}
    \caption{Schematic of the GA. The minimum and maximum considered values for each material parameter are first specified in $\boldsymbol{\theta}_{\rm min} $ and $ \boldsymbol{\theta}_{\rm max}$, respectively. In (1), the binary representation of an initial population $\boldsymbol{\Theta}^{(0)}$ is generated, which consists of $n_{\rm t}$ random material parameter sets $\boldsymbol{\theta}^{(\beta)}$ with $\beta\in\{1,2,\ldots,n_{\rm t}\}$, each within the prescribed range. In (2), the evolution cycle begins, in which $gen\in\{1,2,\ldots,n_{\rm gen}\}$ denotes the generation index. Each material parameter set in the population is passed to the VFM framework to compute the corresponding objective function ${\rm obj}(\boldsymbol{\theta}^{(\beta)})$. In (3), Boltzmann selection is used to select two parent sets, $\boldsymbol{\theta}_{\rm p1}$ and $\boldsymbol{\theta}_{\rm p2}$, from the population $\boldsymbol{\Theta}^{(gen)}$. The probability of being selected depends on the objective function value computed for each material parameter set ${\rm obj}(\boldsymbol{\theta}^{(\beta)})$. The selected parents are then passed into (4): single point crossover, in which a random bit-wise operation is performed on the parent sets to produce two child material parameter sets, $\boldsymbol{\theta}_{\rm c1}$ and $\boldsymbol{\theta}_{\rm c2}$. The two children are passed to (5): bit flip mutation. In this step, a random bit-wise operation is applied to alter the parameters, resulting in the mutated children, $\boldsymbol{\theta}_{\rm cm1}$ and $\boldsymbol{\theta}_{\rm cm2}$. The two mutated child sets are assigned to the next generation's population $\boldsymbol{\Theta}^{(gen+1)}$ in (6). Steps (3)-(6) are repeated until $n_{\rm t}$ mutated child sets are produced, completing $\boldsymbol{\Theta}^{(gen+1)}$. The generation then ends, overwriting $\boldsymbol{\Theta}^{(gen)}$ with $\boldsymbol{\Theta}^{(gen+1)}$, and the GA repeats the evolution cycle in steps (2)-(6). Between two consecutive generations, an elitism strategy is applied to prevent divergence. When $gen$ reaches the target number of generations $n_{\rm gen}$, the material parameter set corresponding to the minimum objective function value in the population is saved in (7). Steps (1)-(7) complete the process for one independent population. Our implementation involves $n_{\rm pop}$ independent initial populations that perform steps (1)-(7) in parallel. In the end, the GA outputs the best performing material parameter set found across all independent populations.} 
    \label{fig:ga_schematic}
\end{figure}

As illustrated in Fig.~\ref{fig:ga_schematic}, the major parts of the GA are the initialization of a population, objective function computation, selection, crossover, and mutation. The steps of objective function computation, selection, crossover, and mutation serve as one evolution cycle, i.e., a generation with index denoted by $gen$. The selection, crossover, and mutation steps combine to mimic the ideas behind natural selection. The evolution cycle terminates when the index $gen$ reaches the specified target number of generations $n_{\rm gen}$. The optimized material parameter set is identified as the set with the lowest objective function value in the population in the final generation. The following sections explain the details of each step in the GA implemented in our framework. 

\subsubsection{Generation of an initial population within a prescribed range} \label{sec:initial_pop}
As illustrated in Fig.~\ref{fig:ga_schematic}(1), the first step in initializing the GA is to prescribe a range for the optimization variables, i.e., the material parameter set for the target constitutive model in our framework. Accordingly, $\boldsymbol{\theta}_{\rm min}$ and $\boldsymbol{\theta}_{\rm max}$ represent sets of the respective minimum and maximum considered values for each material parameter. Then, a population of $n_{\rm t}$ independent material parameter sets $\boldsymbol{\theta}^{(\beta)}$ with $\beta\in\{1,2,\ldots,n_{\rm t}\}$ may be generated. Each candidate $\beta$ must be consistent with $\boldsymbol{\theta}_{\rm min} \le \boldsymbol{\theta}^{(\beta)} \le \boldsymbol{\theta}_{\rm max}$ in an element-by-element sense. To this end, we adopt the common approach of binary encoding of the optimization variables, implemented by discretizing the respective material parameter ranges and mapping the discretized values onto a bit space of binary numbers. Each individual material parameter in $\boldsymbol{\theta}^{(\beta)}$ is represented using $n_{\rm b}$ bits. For example, if $n_{\rm b}=4$ (common usage throughout this study), the four bits have 16 unique combinations. During the mapping between continuous parameter space and discrete bit space, the binary number $0000$ maps onto the respective element of $\boldsymbol{\theta}_{\rm min}$, $1111$ maps onto the respective element of $\boldsymbol{\theta}_{\rm max}$, and the other unique binary numbers in bit space sample within each material parameter range with equal intervals. Each candidate set $\boldsymbol{\theta}^{(\beta)}$ is generated directly in bit space, and the initial population containing $n_{\rm t}$ independent candidate material parameter sets is denoted as $\boldsymbol{\Theta}^{(0)} = \{\boldsymbol{\theta}^{(1)},\boldsymbol{\theta}^{(2)},\ldots,\boldsymbol{\theta}^{(n_{\rm t})}\}$. The superscript on $\boldsymbol{\Theta}^{(0)}$ is the generation index with $gen=0$ denoting the initial population. Since each material parameter set $\boldsymbol{\theta}^{(\beta)}$ contains $n_{\rm param}$ number of parameters, optimization takes place over a scope of $({2^{n_{\rm b}}})^{n_{\rm param}}$ possible combinations of material parameters. For data structure purposes, the population $\boldsymbol{\Theta}^{(0)}$ is arranged in a matrix of bits with $n_{\rm t}$ rows, each representing an independent material parameter set $\boldsymbol{\theta}^{(\beta)}$, and $n_{\rm b} n_{\rm param}$ columns with every $n_{\rm b}$ entries representing a material parameter.

\subsubsection{Computation of the objective function} 
After the initial population is generated, the evolution cycle begins. Referring to Fig.~\ref{fig:ga_schematic}(2), each cycle starts with evaluating the performance of each candidate set $\boldsymbol{\theta}^{(\beta)}$ in the current population $\boldsymbol{\Theta}^{(gen)}$ by computing the respective objective function. For each $\beta\in \{1,2,\ldots,n_{\rm t}\}$, $\boldsymbol{\theta}^{(\beta)}$ is extracted from the respective row in the population matrix $\boldsymbol{\Theta}^{(gen)}$ and converted from discrete bit space to natural numbers. Each converted $\boldsymbol{\theta}^{(\beta)}$ is then passed into equation \eqref{obj} in the VFM module to compute the corresponding objective function value ${\rm obj}(\boldsymbol{\theta}^{(\beta)})$. These computed objective function values are stored in a vector of length $n_{\rm t}$, denoted as $\mathbb{O}^{(gen)}$ with each entry in $\mathbb{O}^{(gen)}$ corresponding to the respective row in $\boldsymbol{\Theta}^{(gen)}$. 

\subsubsection{Boltzmann selection}\label{sec:boltz} 
Once $\mathbb{O}^{(gen)}$ has been calculated, the process of generating the population for the next generation $\boldsymbol{\Theta}^{(gen+1)}$ may begin. To do so, pairs of ``parent'' material parameter sets are selected from generation $gen$ to generate two ``children'' in generation $gen+1$. The first step in this process is selecting pairs of parent sets from the population $\boldsymbol{\Theta}^{(gen)}$, which are then denoted by $\boldsymbol{\theta}_{{\rm p}1}$ and $\boldsymbol{\theta}_{{\rm p}2}$. Since a candidate set $\boldsymbol{\theta}^{(\beta)}$ that better satisfies the balances of internal forces \eqref{balance of internal force 2} and external forces \eqref{balance of external force 3} corresponds to a lower value of the objective function, the selection probability for set $\beta$ depends on ${\rm obj}(\boldsymbol{\theta}^{(\beta)})$, so that parent sets with lower objective function values are more likely to be selected. As illustrated in Fig.~\ref{fig:ga_schematic}(3), the selection mechanism in our implementation is Boltzmann selection \citep[e.g.,][]{goldberg1990note,maza1993analysis,mahfoud2000boltzmann}, which selects parent sets based on their objective function values treated by a Boltzmann kernel with a generation-dependent quasi-temperature $T(gen)$. In Boltzmann selection, the probability of a candidate set $\beta$ being selected is proportional to the quantity $\exp[-T(gen)/{\rm obj}(\boldsymbol{\theta}^{(\beta)})]$. The quasi-temperature $T(gen)$ is defined as a function of generation index $gen$, and its dependence is constructed to decrease linearly from a maximum value $T_{\rm max}$ to a minimum value $T_{\rm min}$ as the generation index increases. With $n_{\rm gen}$ the total number of generations, the functional form for $T(gen)$ is taken to be
\begin{equation}\label{eq:boltz}
 T(gen)=\left\{
  \begin{array}{@{}ll@{}}
    T_{\rm max} & \text{if } gen \le 2, \\
    T_{\rm max} - (T_{\rm max}-T_{\rm min})\dfrac{gen-2}{n_{\rm gen}-4} & \text{if } 2 < gen \le n_{\rm gen}-2, \\
    T_{\rm min} & \text{if } n_{\rm gen}-2 < gen \le n_{\rm gen}. \\
  \end{array}\right.
\end{equation} 
In \eqref{eq:boltz}, $T_{\rm max}$ and $T_{\rm min}$ are two GA hyperparameters. Our implementation of Boltzmann selection is summarized in Algorithm~\ref{BS}. Each element of the vector storing all the objective function values $\mathbb{O}^{(gen)}$ is first treated by the Boltzmann kernel (Lines 1-3) and then normalized by the sum of the resulting components (Line 4) to obtain the individual probability vector $\mathbb{P}^{(gen)}_{\rm i}$, each component of which represents the probability that the respective material parameter set will be selected as a parent set. Lines 5-8 then convert $\mathbb{P}^{(gen)}_{\rm i}$ to a cumulative probability vector $\mathbb{P}^{(gen)}_{\rm c}$. In Lines 9-16, to select a parent set, a random number $r_1$ between zero and one is generated, and the cumulative probability vector is used to determine the corresponding parent set index. Two parent sets are selected in this way and assigned to $\boldsymbol{\theta}_{{\rm p}1}$ and $\boldsymbol{\theta}_{{\rm p}2}$ from $\boldsymbol{\Theta}^{(gen)}$ in lines 17-18. 

\begin{algorithm}[t]
  \caption{Boltzmann Selection}\label{BS}
  \begin{algorithmic}[1]
    \For{$\beta=1,n_{\rm t}$}
    \State $\mathbb{P}^{(gen)}_{\rm i}(\beta) \gets \exp\left[-T(gen)/ {\rm obj}(\boldsymbol{\theta}^{(\beta)})\right]$
    \EndFor
    \State $\mathbb{P}^{(gen)}_{\rm i} \gets \mathbb{P}^{(gen)}_{\rm i}/\sum_{\beta=1}^{n_{\rm t}}\mathbb{P}^{(gen)}_{\rm i}(\beta)$\Comment{Individual probability vector}
    \State $\mathbb{P}^{(gen)}_{\rm c}(1) \gets \mathbb{P}^{(gen)}_{\rm i}(1)$
    \For{$\beta=2,n_{\rm t}$}
        \State $\mathbb{P}^{(gen)}_{\rm c}(\beta) \gets \mathbb{P}^{(gen)}_{\rm c}(\beta-1)+\mathbb{P}^{(gen)}_{\rm i}(\beta)$\Comment{Cumulative probability vector}
    \EndFor
    \For{$i=1,2$}
        \State $r_1 \gets \text{random}(0,1)$
        \State $j \gets 1$
        \While{ $\mathbb{P}^{(gen)}_{\rm c}(j) < r_1$}
            \State $j \gets j+1$
        \EndWhile
        \State $select(i) \gets j$
    \EndFor
    \State $\boldsymbol{\theta}_{{\rm p}1} \gets \boldsymbol{\Theta}^{(gen)}(select(1),:)$
    \State $\boldsymbol{\theta}_{{\rm p}2} \gets \boldsymbol{\Theta}^{(gen)}(select(2),:)$
\end{algorithmic}
\end{algorithm}

We compared Boltzmann Selection to the most common selection mechanism---Roulette Wheel Selection---in which the probability of a candidate set being selected is linearly proportional to the inverse of the objective function value. In contrast, Boltzmann selection enables a more diverse selection in early generations when the quasi-temperature $T(gen)$ is high, while enforcing a more conservative selection in later generations when the quasi-temperature is lower that preserves the best-performing candidate sets. In our experience, we observed that Boltzmann selection is significantly more powerful than Roulette Wheel Selection in determining an optimized material parameter set.

\subsubsection{Single point crossover}\label{sec:crossover} 
In crossover, the two selected parent sets, $\boldsymbol{\theta}_{{\rm p}1}$ and $\boldsymbol{\theta}_{{\rm p}2}$, combine to produce two child sets, $\boldsymbol{\theta}_{{\rm c}1}$ and $\boldsymbol{\theta}_{{\rm c}2}$, as shown in Fig.~\ref{fig:ga_schematic}(4). Reproduction takes place in bit space by switching certain binary numbers at the same indices in $\boldsymbol{\theta}_{{\rm p}1}$ and $\boldsymbol{\theta}_{{\rm p}2}$. When a subset of bits are switched, the child sets possess different material parameters than their parent sets and hence correspond to different values of the objective function. While the selected parent sets typically exhibit strong performance quantified by the objection function, a child set that outperforms its parent sets is likely to emerge. 

A number of crossover operators have been previously developed in the literature \citep{larranaga1999genetic,lambora2019genetic,katoch2021review}, e.g., Single Point Crossover, Two Point Crossover, Uniform Crossover, and Shuffle Crossover. Here, we utilize Single Point Crossover due to its simplicity and efficiency. Our implementation of Single Point Crossover is summarized in Algorithm \ref{SX}. A random number $r_1$ between zero and one is generated in line 1. Line 2 checks $r_1$ against a hyperparameter $0<x_{\rm rate}<1$, if $r_1 \ge x_{\rm rate}$, crossover does not occur, and the parent sets, $\boldsymbol{\theta}_{{\rm p}1}$ and $\boldsymbol{\theta}_{{\rm p}2}$, are copied to the child sets, $\boldsymbol{\theta}_{{\rm c}1}$ and $\boldsymbol{\theta}_{{\rm c}2}$. If $r_1<x_{\rm rate}$ (Line 5), the crossover event takes place, and in Lines 6--18 each material parameter in the set is individually considered. For a given material parameter, in Line 7, a second random number $r_2$ between zero and one is generated. Like $r_1$, it is compared against a hyperparameter $x_{\rm con}$ to determine if crossover for that parameter occurs. In Lines 8--10, if $r_2 \ge x_{\rm con}$, crossover does not occur, and the first and second child sets possess the same values of that material parameter as the first and second parent sets, respectively. In Lines 11--16, if $r_2 < x_{\rm con}$, Single Point Crossover takes place for that parameter. A random integer between 1 and $n_{\rm b}-1$ is generated as the crossover site index in bit space, and the bits after the site index $r_3$ in the binary numbers corresponding to the parameter in the two parent sets are switched, producing two new values of the material parameter in bit space. These values are then assigned to the respective child sets. In sum, Single Point Crossover involves two hyperparameters, $x_{\rm rate}$ and $x_{\rm con}$, with $x_{\rm rate}$ controlling the probability of crossover occurring on the set level and $x_{\rm con}$ controlling the probability of crossover occurring on the individual material parameter level. 

\begin{algorithm}[t]
\caption{Single Point Crossover}\label{SX}
\begin{algorithmic}[1]
\State $r_1 \gets \texttt{random}(0,1)$
\If{$r_1 \ge x_{\rm rate}$}
    \State $\boldsymbol{\theta}_{{\rm c}1} \gets \boldsymbol{\theta}_{{\rm p}1}$
    \State $\boldsymbol{\theta}_{{\rm c}1} \gets \boldsymbol{\theta}_{{\rm p}2}$
\ElsIf{$r_1<x_{\rm rate}$}
    \For{$i=1,n_{\rm param}$}
        \State $r_2 \gets \texttt{random}(0,1)$
        \If{$r_2 \ge x_{\rm con}$}
            \State $\boldsymbol{\theta}_{{\rm c}1}((1+n_{\rm b}(i-1)):(n_{\rm b}i)) \gets \boldsymbol{\theta}_{{\rm p}1}((1+n_{\rm b}(i-1)):(n_{\rm b}i))$
            \State $\boldsymbol{\theta}_{{\rm c}2}((1+n_{\rm b}(i-1)):(n_{\rm b}i)) \gets \boldsymbol{\theta}_{{\rm p}2}((1+n_{\rm b}(i-1)):(n_{\rm b}i))$
        \ElsIf{$r_2<x_{\rm con}$}
            \State $r_3 \gets \texttt{random\_int}(1,(n_{\rm b}-1))$
            \State $\boldsymbol{\theta}_{{\rm c}1}((1+n_{\rm b}(i-1)):(r_3+n_{\rm b}(i-1))) \gets \boldsymbol{\theta}_{{\rm p}1}((1+n_{\rm b}(i-1)):(r_3+n_{\rm b}(i-1)))$
            \State $\boldsymbol{\theta}_{{\rm c}2}((1+n_{\rm b}(i-1)):(r_3+n_{\rm b}(i-1))) \gets \boldsymbol{\theta}_{{\rm p}2}((1+n_{\rm b}(i-1)):(r_3+n_{\rm b}(i-1)))$
            \State $\boldsymbol{\theta}_{{\rm c}1}((r_3+1+n_{\rm b}(i-1)):(n_{\rm b}i)) \gets \boldsymbol{\theta}_{{\rm p}2}((r_3+1+n_{\rm b}(i-1)):(n_{\rm b}i))$
            \State $\boldsymbol{\theta}_{{\rm c}2}((r_3+1+n_{\rm b}(i-1)):(n_{\rm b}i)) \gets \boldsymbol{\theta}_{{\rm p}1}((r_3+1+n_{\rm b}(i-1)):(n_{\rm b}i))$
        \EndIf
    \EndFor
\EndIf
\end{algorithmic}
\end{algorithm}

\subsubsection{Bit flip mutation}\label{sec:mutation} 
The child sets, $\boldsymbol{\theta}_{{\rm c}1}$ and $\boldsymbol{\theta}_{{\rm c}2}$, are then subjected to a mutation operation (Fig.~\ref{fig:ga_schematic}(5)), in which certain binary numbers in the child sets may be reversed. Mutation adds randomness to the population, which helps maintain diversity and enables the GA to escape local minima. There are a number of mutation operators in the literature \citep{larranaga1999genetic,lambora2019genetic,katoch2021review}, e.g., Bit Flip Mutation, Swap Mutation, and Gaussian Mutation. Here, we utilize Bit Flip Mutation due to its fine-grained variability and simplicity. The process, which operates on one child set $\boldsymbol{\theta}_{{\rm c}}$ at a time, is summarized in Algorithm \ref{BF}. Like Single Point Crossover, Bit Flip Mutation involves two hyperparameters, $m_{\rm rate}$ and $m_{\rm con}$, which control the probability of mutation occurring on the set level and the individual material parameter level, respectively. If mutation happens for an individual parameter, a random integer $r_3$ ranging from 1 to $n_{\rm b}$ indicates the index at which the bit is either flipped from 0 to 1 or from 1 to 0. After mutation, the unmutated child set $\boldsymbol{\theta}_{{\rm c}}$ is replaced by the mutated child set $\boldsymbol{\theta}_{{\rm cm}}$.

\begin{algorithm}[t]
\caption{Bit Flip Mutation}\label{BF}
\begin{algorithmic}[1]
\State $r_1 \gets \texttt{random}(0,1)$
\If{$r_1 \ge m_{\rm rate}$}
    \State $\boldsymbol{\theta}_{\rm cm} \gets \boldsymbol{\theta}_{\rm c}$
\ElsIf{$r_1<m_{\rm rate}$}
    \For{$i=1,n_{\rm param}$}
        \State $\boldsymbol{\theta}_{\rm cm}((1+n_{\rm b}(i-1)):(n_{\rm b}i)) \gets \boldsymbol{\theta}_{\rm c}((1+n_{\rm b}(i-1)):(n_{\rm b}i))$
        \State $r_2 \gets \texttt{random}(0,1)$
        \If{$r_2 < m_{\rm con}$}
            \State $r_3 \gets \texttt{random\_int}(1,n_{\rm b})$
            \State $\boldsymbol{\theta}_{\rm cm}(r_3+n_{\rm b}(i-1)) \gets \,\sim\boldsymbol{\theta}_{\rm cm}(r_3+n_{\rm b}(i-1))$
        \EndIf
    \EndFor
\EndIf
\end{algorithmic}
\end{algorithm}

\subsubsection{Termination condition for one generation and one population} 
Each pair of mutated child sets, $\boldsymbol{\theta}_{{\rm cm}1}$ and $\boldsymbol{\theta}_{{\rm cm}2}$, created through the selection-crossover-mutation process, is then stored in the matrix $\boldsymbol{\Theta}^{(gen+1)}$ (Fig.~\ref{fig:ga_schematic}(6)), which represents the next generation of the population. With $k$ denoting the pair-selection index, the selection-crossover-mutation-update loop (Fig.~\ref{fig:ga_schematic}(3)-(6)) repeats until $k$ reaches $n_{\rm t}/2$, filling $\boldsymbol{\Theta}^{(gen+1)}$ with $n_{\rm t}$ mutated offspring. The current generation then terminates, and the GA moves on to the next generation, repeating steps (2)-(6) as indicated in Fig.~\ref{fig:ga_schematic}. The GA stops producing new generations when the target number of generations $n_{\rm gen}$ is reached. At this point, all members of the final generation $\boldsymbol{\Theta}^{(n_{\rm gen})}$ are evaluated by computing the corresponding objective function values ${\mathbb O}^{(n_{\rm gen})}$. The material parameter set corresponding to $\min({\mathbb O}^{(n_{\rm gen})})$ is then saved (Fig.~\ref{fig:ga_schematic}(7)). 

\subsubsection{Convergence} \label{sec:strat}
The convergence of the GA in the context of our framework is assessed by the improvement in $\min({\mathbb O}^{({gen})})$ over the generations for a certain initial population. Convergence is attained when the improvement in $\min({\mathbb O}^{({gen})})$ is sufficiently small over a certain number of generations. However, there are two issues in GA convergence that require further attention.

\paragraph{Divergence and elitism} The first issue is the possibility of divergence. The evolution cycle is designed to strive for, but cannot guarantee, improvement in the minimum objective function value over the generations due to the inherent stochastic nature of the selection-crossover-mutation process. This uncertainty is particularly pronounced as $gen$ approaches $n_{\rm gen}$, and the rate of convergence becomes comparatively slower. To address this point, we implement Elitism \cite[e.g.,][]{bandyopadhyay1995pattern,bhandari1996genetic,ahn2003elitism}, which is a memory-based strategy in GA to prevent divergence. Elitism works by identifying and temporarily storing $\min({\mathbb O}^{({gen})})$ and the corresponding best-performing material parameter set, denoted as $\boldsymbol{\theta}_{\rm best}^{(gen)}$, for a generation $gen$. Then, the GA progresses to the next generation $gen+1$, generating $\boldsymbol{\Theta}^{(gen+1)}$ and computing the corresponding objective function values ${\mathbb O}^{({gen+1})}$. If $\min({\mathbb O}^{({gen+1})})<\min({\mathbb O}^{({gen})})$, then the GA has made progress towards convergence, and the best-performing material parameter set $\boldsymbol{\theta}_{\rm best}^{(gen+1)}$ is updated accordingly. If $\min({\mathbb O}^{({gen+1})})>\min({\mathbb O}^{({gen})})$, divergence is detected, and Elitism activates by first identifying the worst-performing material parameter set in generation $gen+1$, $\boldsymbol{\theta}_{\rm worst}^{(gen+1)}$, corresponding to $\max({\mathbb O}^{({gen+1})})$. Elitism then overwrites $\boldsymbol{\theta}_{\rm worst}^{(gen+1)}$ with $\boldsymbol{\theta}_{\rm best}^{(gen)}$ in the matrix $\boldsymbol{\Theta}^{(gen+1)}$, so that $\boldsymbol{\theta}_{\rm best}^{(gen+1)}=\boldsymbol{\theta}_{\rm best}^{(gen)}$, and ${\mathbb O}^{({gen+1})}$ is updated accordingly. In this way, Elitism unequivocally precludes divergence in $\min({\mathbb O}^{({gen})})$ as the generations progress.

\paragraph{Multipopulation GA} The second issue is the possibility of convergence of the GA to a local minimum for a given initial population. This possibility stems from the fact that a single initial population contains a limited number of candidate sets, only representing a small subset of the possible combinations of material parameters within their prescribed ranges. Although mutations encourage the GA to venture beyond local minima, there remains a possibility for a single initial population to gravitate towards a local minimum. One remedy is to utilize a large population size $n_{\rm t}$; however, this approach is computationally costly. Instead, we adopt a strategy of utilizing multiple initial populations, i.e., multipopulation GA \citep[e.g.,][]{cantu1999topologies}. Our implementation involves a large number of initial populations $n_{\rm pop}$, while maintaining a relatively modest number of candidate sets $n_{\rm t}$ within each population, without communication (e.g., migration) between the populations. From our empirical experience, increasing the number of initial populations significantly improves the likelihood of identifying a material parameter set that approximates the global minimum, compared to utilizing a single large population at similar computational cost. This strategy also has a practical advantage. Because the initial populations are independent of one other, the framework may then be straightforwardly parallelized by distributing numerous independent populations to different CPU threads \citep[e.g.,][]{gorges1990explicit} with each independent initial population undergoing its own GA process without any communication with other populations. Here, we adhere to a guideline of taking $n_{\rm pop} \ge 400$ and $n_{\rm t} \le 1000$ throughout. When the number of completed independent populations reaches $n_{\rm pop}$, the whole GA terminates and gathers the best candidate in each of the populations. The candidate material parameter set that corresponds to the lowest objective function value is selected as the output calibrated material parameter set. 

\begin{figure}[!t]
    \centering
    \includegraphics[width=0.5\textwidth]{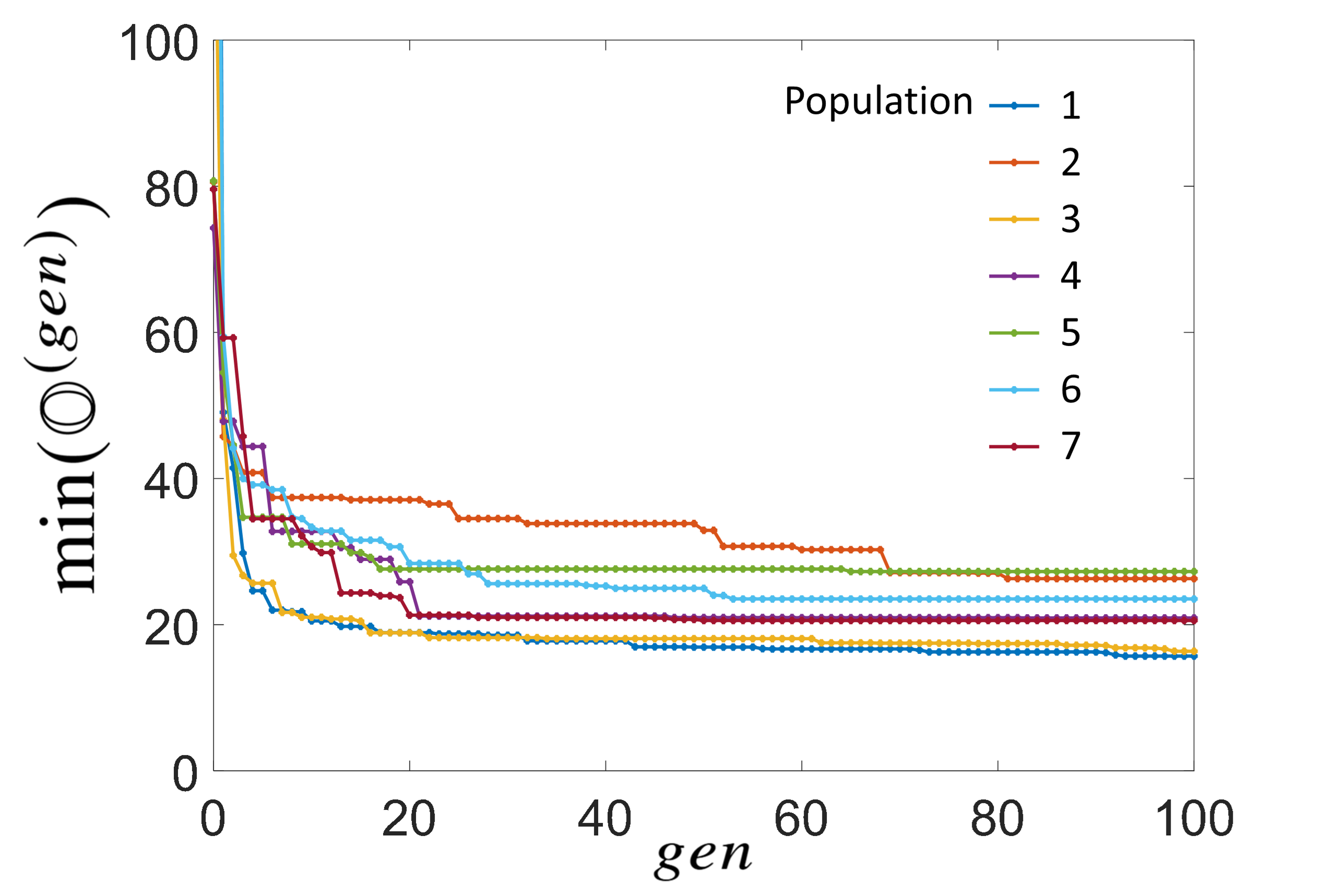}
    \caption{Example of the convergence behavior of the GA for seven independent initial populations selected from $n_{\rm pop} = 500$ populations from the application of VFM-GA to calibrating high-density Poron XRD foam in Section \ref{sec:1st_1}. Convergence is characterized using the population-wise minimum objective function $\min({\mathbb O}^{({gen})})$ versus the generation index $gen$ in the evolution cycles of the GA.}
    \label{fig:conv}
\end{figure}

\paragraph{Convergence example} Here, we provide a representative example of the convergence behavior of the GA, specifically, for the calibration of the high-density Poron XRD foam described in Section~\ref{sec:1st_1}. In Fig.~\ref{fig:conv}, the minimum objective function $\min({\mathbb O}^{({gen})})$ of seven independent populations, chosen from among $n_{\rm pop} = 500$ independent populations, is plotted against the generation index $gen$. Population 1 is the population that leads to the final calibrated material parameter set given in Table~\ref{tab:1st_paras} and applied in Fig.~\ref{fig:1st_curves}(c), and Populations 2 through 7 are selected at random for comparison. One characteristic to notice is that all the population-wise minimum objective functions $\min({\mathbb O}^{({gen})})$ quickly converge as $gen \le 20$. Subsequently, the rate of convergence becomes more gradual when $gen > 20$, and convergence is reached as $n_{\rm gen}-gen \lesssim 10$ (where $n_{\rm gen}=100$ here). Other applications of the VFM-GA framework using different hyperparameters can lead to slightly different convergence behavior as the generation index $gen$ increases, but the general trend of convergence is observed to be the same. Another feature to note is the stepwise improvement of $\min({\mathbb O}^{({gen})})$ that occurs after the initial rapid convergence, which is particularly evident in Population 2. The stepwise improvements possibly result from mutations within the GA's evolution cycle enabling the GA to escape local minima. It is also worth observing that each population's $\min({\mathbb O}^{({gen})})$ converges to its own unique value. Only by utilizing a sufficient number of independent populations $n_{\rm pop}$, as mentioned in the previous paragraph, can one claim that the GA effectively endeavors to approximate the global minima. Finally, it is crucial to note the fact that $\min({\mathbb O}^{({gen})})$ for any population never converges to 0 in this prototypical example. The main reason is the inability of an idealized constitutive model to perfectly capture experimental data. That said, the calibrated result from Population 1 outperforms prior manual calibration \citep{landauer2019experimental}, as illustrated in Fig.~\ref{fig:1st_curves}(c) and discussed further in Section~\ref{sec:1st_1}. Besides the imperfect nature of the constitutive model, two minor factors that hinder convergence to zero are (1) the discretization of the material parameter ranges into bit space, as described in Section \ref{sec:initial_pop}, and (2) the plane-stress approximation, described Section \ref{sec:stress_field}. 

\section{Implementation variants: Two functionalities}\label{sec:implementations}
With the VFM-based objective function detailed in Section \ref{sec:VFM} and our GA methodology detailed in Section \ref{sec:GA}, we may proceed to applying the VFM-GA calibration framework in a code package. To accommodate experimental data from different types of experiments, we pursue two functionalities. The first functionality, abbreviated as ``1st-func'', is specialized for homogeneous deformation experiments, such as simple compression and/or tension, and utilizes engineering stress-strain and engineering lateral-axial strain curves as the experimental input. The second functionality, abbreviated as ``2nd-func'', utilizes data from experiments with inhomogeneous deformation, namely, full-field displacement fields and synchronized load cell data, as described in Section \ref{sec:VFM}.

\subsection{First functionality (1st-func)}\label{sec:1st_func_overview}
\begin{figure}[!t]
    \centering
    \includegraphics[width=1\textwidth]{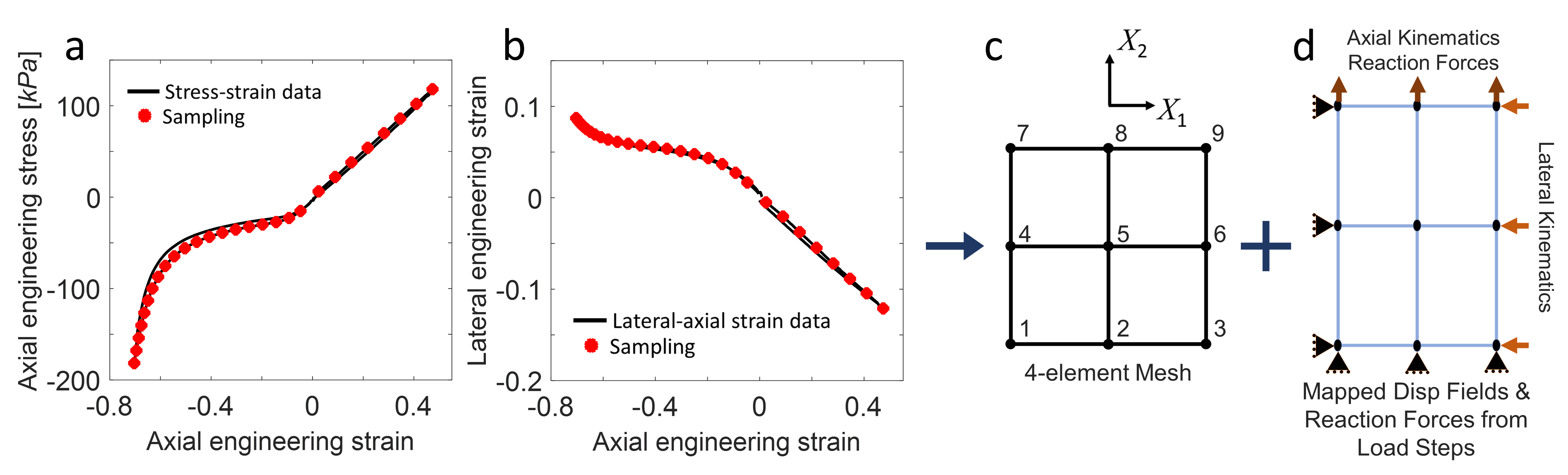}
    \caption{Data conversion preprocessor for the 1st-func. (a) Engineering stress-strain curve with experimental data for the high density Poron XRD foam shown by the solid black curve and the sampling load steps indicated by the red dots. The increment between two sampling load steps along the paths of compression and tension are defined by an equal arclength along the respective engineering stress-strain curves. The sampling points are mapped onto (b) the engineering lateral-axial strain curves at the same sampled axial strain values. The axial and lateral kinematics and the stress at each load step are projected onto (c) a four-element undeformed mesh consisting of quadrilateral elements, where (d) the reaction forces at the boundaries and the displacement field are computed assuming homogeneous deformation.}
    \label{fig:first_func}
\end{figure}

The 1st-func of the VFM-GA framework takes in engineering stress-strain and engineering lateral-axial strain curves as the experimental input data. These two curves may be obtained in simple compression and/or tension experiments, providing kinematics and reaction force information along the load path. This information is then projected onto a uniform mesh, using a data conversion preprocessor, to provide compatibility with the VFM framework. Figure~\ref{fig:first_func} illustrates the workflow of the data conversion preprocessor for the 1st-func with the engineering stress-strain curve shown in Fig.~\ref{fig:first_func}(a) and the engineering lateral-axial strain curve shown in Fig.~\ref{fig:first_func}(b). Sampling of load steps along the stress-strain curve is performed separately for compression and tension. Two parameters that must be specified are $n_{\rm comp}$, the number of load steps sampled in compression, and $n_{\rm ten}$, the number of load steps sampled in tension, so that $n_{\rm comp}+n_{\rm ten} = n_{\rm ls}$. These parameters should be carefully selected based on the experimental dataset and the constitutive model under consideration. For example, selecting $n_{\rm comp}$ to be larger than $n_{\rm ten}$ naturally places a greater weight on fitting the compression response in the calibration process.

Beyond the number of load steps, the manner in which load steps are sampled is important, especially for materials with a highly nonlinear response. For example, for the elastomeric foam material under consideration here, the tangent modulus in compression is low in the plateau regime but high in the densification regime. Sampling load steps with an equal strain increment would result in more load steps being sampled in the plateau regime, placing an unintentional weight on the plateau regime in the fitting process. Conversely, sampling load steps based on an equal stress increment would result in more load steps being sampled in the densification regime. Our strategy to avoid biasing the fit to certain parts of the stress-strain response is to sample load steps based on the arclength along the engineering stress-strain curve. In short, the arclength along the engineering stress-strain curve is numerically integrated using the experimental data in compression and tension separately, and each total arclength is divided by $n_{\rm comp}$ and $n_{\rm ten}$, respectively, to determine the engineering stress and strain for each load step. After the engineering stress and strain values are computed for all sampling load steps, the lateral-axial strain curve is sampled using the axial strain values of the sampling load steps from the engineering stress-strain curve. The result of the sampling procedure is illustrated in Figs.~\ref{fig:first_func}(a) and (b), using $n_{\rm comp}=20$ and $n_{\rm ten}=8$, for the engineering stress-strain and engineering lateral-axial strain curves, respectively, for the high density Poron XRD foam, in which the sampling points are denoted by red dots. We note that, in Figs.~\ref{fig:first_func}(a) and (b), the unloading paths are slightly different compared to the loading paths. For all of the experimental data considered in the present work, the hysteresis is sufficiently small that we may idealize the material as hyperelastic. Accordingly, when applying the 1st-func data conversion preprocessor, sampling is done only on the loading part of the curve. 

Next, we initialize a four-element undeformed mesh, occupying the unit square and consisting of four quadrilateral elements, as illustrated in Fig.~\ref{fig:first_func}(c). In Fig.~\ref{fig:first_func}(d), we apply the axial strain and lateral strain information for each sampled load step onto the undeformed mesh to obtain a displacement field that can be processed by the VFM framework of Section \ref{sec:VFM}. The reaction forces acting on the top and bottom surfaces are obtained through the engineering axial stress times the unit width of the undeformed mesh for each sampled load step. After projection of every sampled load step, we have the necessary information for the VFM framework. We note that, rather than converting the engineering stress-strain and lateral-axial strain data to be VFM-GA compatible, one could alternatively formulate a new objective function explicitly defined through the input and resultant curves, which is a common approach in the past literature on material calibration using GA \citep[e.g.,][]{furukawa1997inelastic,fernandez2018genetic}. If one is fitting only engineering stress-strain data, this is a straightforward approach, but, to collectively fit both engineering stress-strain and lateral-axial strain data, one would need to define residuals for each curve and then combine them into one objective function. The objective function \eqref{obj} already offers a systematic approach for balancing these contributions. The axial stress-strain response and the lateral-axial strain response are accounted for by the balance of external forces \eqref{balance of external force 3} and internal forces \eqref{balance of internal force 2}, respectively, which are then combined and normalized in the objective function \eqref{obj}. Therefore, the objective function \eqref{obj} may be straightforwardly leveraged for calibration using homogeneous deformation experiments.

\subsection{Second functionality (2nd-func)}\label{sec:2nd_func_overview}
The 2nd-func of the VFM-GA framework is the native implementation that takes in full-field displacement fields and synchronized load cell data as input from experiments. It is compatible with both homogeneous and inhomogeneous displacement fields and may be applied to one experiment at a time or multiple experiments collectively. For the latter, multiple undeformed meshes are generated, each having their own corresponding displacement fields and synchronized load cell data. To calculate the objective function, the VFM framework processes each experiment separately, using the workflow of Fig.~\ref{fig:framework}, and, in step (13), the load steps from each experiment are simply added together in the objective function, exploiting the fact that the contribution of each load step $m$ to the objective function in any experiment is normalized by  $\|\mathbb{F}_{\mathrm{ext}}^{\mathrm{exp}}\|_2^{(m)}$ in \eqref{obj}. For illustration, in Section~\ref{sec:2nd_homo}, the 2nd-func will be collectively applied to data from two homogeneous deformation experiments (simple compression and tension), while in Section~\ref{sec:2nd_inhomo}, the 2nd-func will be applied to data from a single wedge indentation experiment. 

As for the 1st-func, the sampling of load steps is important, but, since the 2nd-func is designed to accommodate experiments with different types of deformation fields, the systematic procedure for sampling load steps described in Section~\ref{sec:1st_func_overview} cannot be uniformly applied. Instead, load steps should be chosen based on the nature of the deformation fields in a given experiment and the features of the target constitutive model. When selecting load steps, one should ensure that the load steps span the intended range of deformation states over which the model is to be fit. Moreover, it is beneficial to bias the selection of load steps towards the more nonlinear regimes of deformation. For instance, the compressible elastomeric foams under consideration here display a more nonlinear response in compression than in tension, so the number of load steps involving compression-dominated deformation should be greater than the number of load steps involving tension-dominated deformation. 

Regarding full-field displacement fields, issues, such as imperfect lubrication, out-of-plane barreling, and imperfect lighting, can limit the accuracy of the displacement fields calculated by a DIC algorithm near the boundary of a specimen on which the displacement is prescribed, and this inaccuracy can be a significant source of error in objective function computation. The remedy applied in this work is to crop a small portion of the undeformed mesh and the corresponding displacement fields near prescribed-displacement boundaries. The cropping must satisfy the condition that the net force due to the traction distribution on the cropping plane equals the net force on the corresponding prescribed-displacement boundary. We note that the STAQ-DIC algorithm used in this work \citep{Yang2022} can accurately compute the displacement fields near free surfaces, so no cropping is applied for this type of boundary. As a guideline, the number of elements removed due to cropping at a prescribed-displacement boundary should be less than 10\% of the total number of elements in the undeformed mesh to ensure that a sufficient portion of the displacement field is retained for the VFM framework. 

\subsection{Code packages}
The 1st-func and 2nd-func of the VFM-GA framework have been implemented in two separate code packages in Fortran 90, which may be compiled using the Intel Fortran Compiler (ifx) from the Intel oneAPI HPC Toolkit \citep{intel_oneapi_hpc_toolkit}. For both packages, three Matlab scripts serve as the interface. The first Matlab script initializes the undeformed mesh and the experimental displacement fields with synchronized load cell data. For the 1st-func, this involves applying the data conversion preprocessor described in Section~\ref{sec:1st_func_overview} and illustrated in Fig.~\ref{fig:first_func}. The second Matlab script specifies the ranges of the material parameters that define the scope of GA optimization (Appendix~\ref{app:mat para}), the VFM parameters (Appendix~\ref{app:code para}), and the GA hyperparameters (Appendix~\ref{app:hyper para}). These inputs are then written to a global Fortran 90 module to be used by the code package. The third MATLAB script supports two execution workflows for the VFM-GA framework. It either compiles the Fortran codes and executes them in parallel on a Windows platform using Visual Basic Scripting, or it prepares a folder containing source files and batch scripts that can be transferred to a Linux-based cluster for compilation and parallel execution. Both code packages are built using the hyperelastic constitutive model described in Section~\ref{sec:model}. Modifying the code packages to be applied to other hyperelastic models requires the modification of a subroutine in the Fortran 90 code, which takes in the deformation gradient $\bfF$ and the set of material parameters $\boldsymbol{\theta}$ and outputs the Kirchhoff stress $J\bfT=\bfP\bfF^\top$. The code packages are available on GitHub at the following link: https://github.com/HenannResearchGroup/VFM-GA.

\section{Application of the VFM-GA framework to experimental data}\label{sec:results}
In this section, we apply both functionalities of the VFM-GA framework to calibrating the hyperelastic constitutive model described in Section \ref{sec:model} to experimental data for a polyurethane-based, open-cell elastomeric foam (Poron XRD, Rogers Corp, Rogers, CT). Poron XRD foam exhibits a nonlinear response under large deformation in compression, characterized by an initial linear regime followed by plateau and densification regimes, and was experimentally characterized and modeled in \citet{landauer2019experimental}. The manual calibration procedure for the constitutive model outlined in \citet{landauer2019experimental} is complex, and therefore, this material and constitutive model serve as an excellent test case for demonstrating the capability of the VFM-GA framework.

\subsection{Application of the 1st-func}\label{sec:1st}
\subsubsection{Inputs for the 1st-func}\label{sec:1st_1}
\paragraph{VFM inputs} For the experimental input to the 1st-func, we use engineering stress-strain curves and engineering lateral-axial strain curves for three densities of Poron XRD foam: 144\,kg/m$^3$ (low density), 192\,kg/m$^3$ (medium density), and 240\,kg/m$^3$ (high density). All of the stress-strain and lateral-axial strain curves originate from simple compression/tension experiments conducted at a constant engineering strain rate of $2\times10^{-4}$\,s$^{-1}$. The experimental data for the low and moderate density foams are from \citet{landauer2019experimental}. New experimental data for the high-density foam was collected for the application of the 2nd-func described in Section~\ref{sec:2nd_homo} and, for consistency, is also used in this section. 

\paragraph{GA inputs} The GA requires the user to prescribe the ranges for each material parameter in the constitutive model. There are 14 parameters in total: $\{ G_{0}$, $B$, $J_{\rm min}$, $C_1$, $K_1^{0}$, $\Delta_K$, $X_1^\prime$, $X_2^\prime$, $C_0$, $p$, $q$, $r$, $C_2$, $C_3\}$. The prescribed ranges for the material parameters are documented table in Tables~\ref{tab:prescribed_12} and \ref{tab:prescribed_gb} in Appendix~\ref{app:mat para}. In this constitutive model, the first two material parameters, $G_0$ and $B$, are the ground-state shear and bulk moduli, respectively. Following \citet{landauer2019experimental}, these parameters may be determined through the Young's modulus and Poisson's ratio obtained from the initial slopes of the stress-strain and lateral-axial strain curves, respectively. Accordingly, these two parameters are fixed to be the same as the manually fitted values in \citet{landauer2019experimental} here, thus reducing the dimension of the optimization problem to 12. For the remaining 12 parameters, which are non-dimensional, their ranges are chosen to be broad enough to include not only the material parameters determined for all densities of the open-cell foams in \citet{landauer2019experimental} but also the parameters determined for a closed-cell foam in \citet{li2022large}. Besides the prescribed ranges of material parameters, the weight parameter, number of load steps, and stability filter parameters are documented in Table \ref{tab:VFM_parameters} of Appendix~\ref{app:code para}, and the GA hyperparameters are documented in Table \ref{tab:GA_hyperparameters} of Appendix~\ref{app:hyper para}.

\subsubsection{Results for the 1st-func}\label{sec:1st_2}
\begin{figure}[!t]
    \centering
    \includegraphics[width=0.9\textwidth]{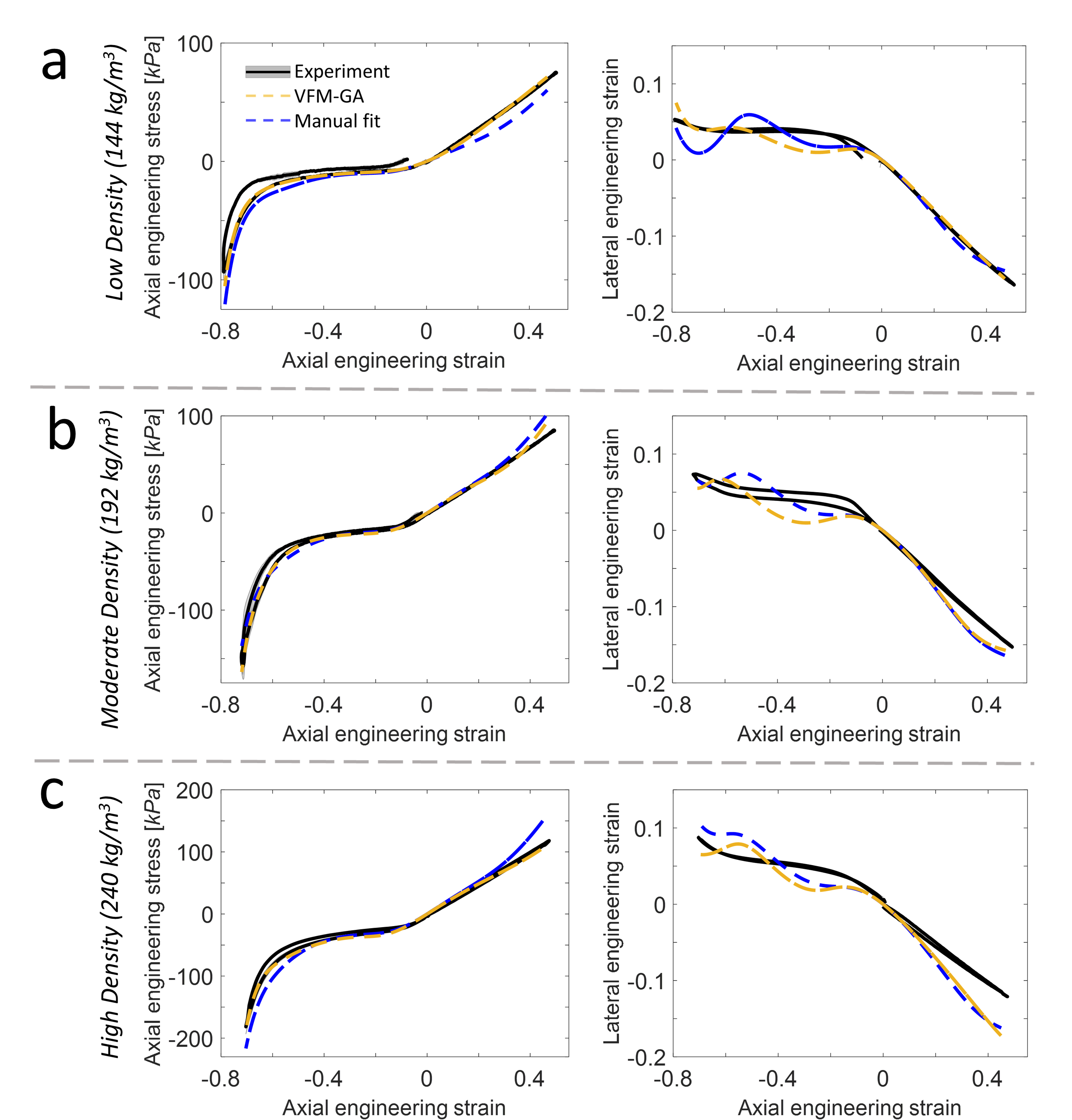}
    \caption{Results for the 1st-func in simple compression and tension. Axial engineering stress versus axial engineering strain (left column) and lateral engineering strain versus axial engineering strain (right column) for Poron XRD foams of (a) low density, (b) medium density, and (c) high density. Solid black lines with shaded error regions (one standard deviation, nine experiments for (a) and (b); one standard deviation, four experiments for (c)) indicate experimental data. Dashed orange lines indicate constitutive model fits using VFM-GA material parameters, and dashed blue lines indicate constitutive model fits using the manually fit material parameters from \citet{landauer2019experimental}.}
    \label{fig:1st_curves}
\end{figure}

The material parameter sets for each density of Poron XRD foam, determined through the application of the VFM-GA framework, are summarized Table \ref{tab:1st_paras}, along with the manually fitted parameter sets determined in \citet{landauer2019experimental} for comparison. Figure \ref{fig:1st_curves} shows the resulting engineering stress-strain curves (left column) and lateral-axial strain curves (right column) in simple compression and tension obtained using the VFM-GA material parameters (dashed orange lines) and the manually fit material parameters (dashed blue lines), along with the experimental data (solid black lines with shaded error regions), for each density of Poron XRD foam. In the stress-strain curves, the model fits using the VFM-GA parameters capture the experimental data in compression accurately, outperforming the manual fit, especially in the plateau and densification regimes (i.e., for compressive engineering strains greater than 0.3). In addition, the VFM-GA fit outperforms the manual fit in the large-deformation tension response. While the lateral-axial strain response is not perfectly captured, mainly due to the design of the constitutive model that prioritizes capturing the stress-strain response at the expense of capturing the details of the lateral response, the VFM-GA fit improves upon the manual fit. We note that for the manual fit in \citet{landauer2019experimental}, the material parameters besides \{$G_0$,$B$,$J_{\rm min}$,$C_1$\} were treated as constant across the different densities of Poron XRD, which was a practical simplification due to the challenge of manual fitting. The VFM-GA framework does not necessitate this simplification, unleashing the potential of the constitutive model by delivering improved fits while minimizing effort on the part of the user.

\begin{table}[!t]
\setlength{\tabcolsep}{3.96pt} 
{\small
\centering
 \begin{tabular}{cccccccccccccccc}
 \toprule
 Density & Fit & $G_0$ [kPa] & $B$ [kPa] & $J_{\rm min}$ & $C_1$ & $K_1^0$ & $\Delta_{\rm K}$ & $X_1^{\prime}$ & $X_2^{\prime}$ & $C_0$ & $p$ & $q$ & $r$ & $C_2$ & $C_3$ \\ 
 \noalign{\smallskip}\hline\noalign{\smallskip}
 \multirow{2}{*}{Low}  & Manual & \multirow{2}{*}{34.5} & \multirow{2}{*}{58.7} & 0.12 & 2.5 & -0.21 & 0.2 & 3.7 & 0.22 & 0.1 & 4 & 5 & 2 & 9 & 0.026 \\
 & VFM-GA & & & 0.18 & 1.24 & -0.0333 & 0.283 & 6.8 & 0.208 & 0.05 & 4.8 & 3.2 & 1 & 13  & 0.134 \\
 \noalign{\smallskip}\hline\noalign{\smallskip}
 \multirow{2}{*}{Medium} & Manual & \multirow{2}{*}{65.2} & \multirow{2}{*}{117.4} & 0.16 & 1.9 & -0.21 & 0.2 & 3.7 & 0.22 & 0.1  & 4 & 5 & 2 & 9 & 0.026 \\
 & VFM-GA & & & 0.247 & 4.02 & -0.2 & 0.307 & 3.6 & 0.076 & 0.05 & 5.6 & 6.4 & 1 & 10.3  & 0.0343 \\
 \noalign{\smallskip}\hline\noalign{\smallskip}
 \multirow{2}{*}{High}& Manual & \multirow{2}{*}{102} & \multirow{2}{*}{193.8} & 0.19 & 1.9 & -0.21 & 0.2 & 3.7 & 0.22 & 0.1  & 4 & 5 & 2 & 9 & 0.026 \\
 & VFM-GA & & & 0.113 & 0.843 & -0.3 & 0.167 & 2.53 & 0.274 & 0.05 & 6.4 & 5.2 & 2.3 & 9  & 0.0343 \\
 \hline
\end{tabular}
}
\caption {Comparison of material parameters for low, moderate, and high density Poron XRD foams determined using the 1st-func of the VFM-GA framework versus manual fitting. In this table, we represent the dimensionless parameters determined by the VFM-GA framework with no more than three leading digits, but this is not meant to reflect precision. In the VFM-GA framework, the precision of a fitted parameter stems from the discretization of its parametric space.} 
\label{tab:1st_paras}
\end{table}

\subsubsection{Validation tests for the 1st-func}\label{sec:1st_3}
\begin{figure}[!t]
    \centering
    \includegraphics[width=1\textwidth]{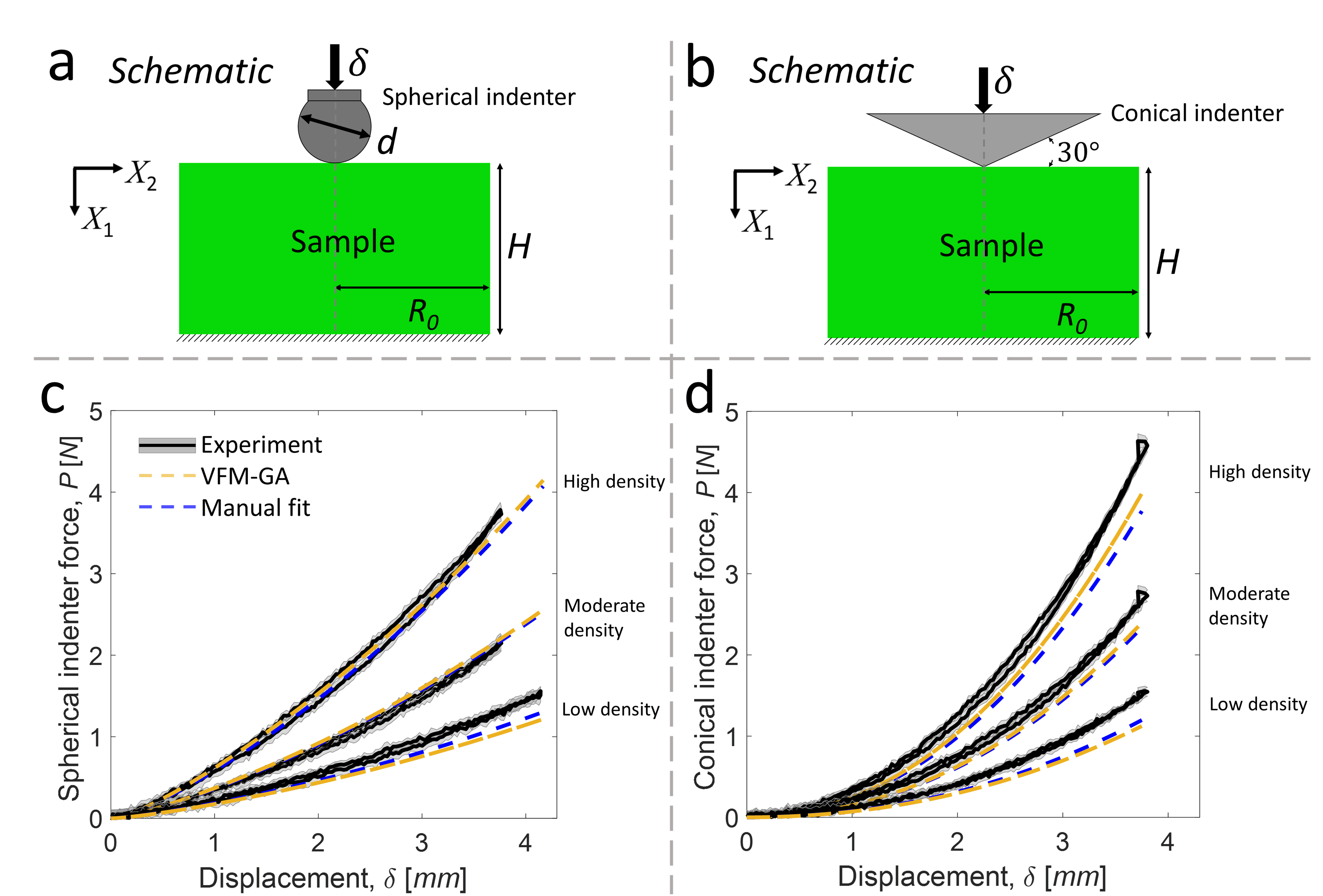}
    \caption{Validation test for the 1st-func: Spherical and conical indentation. Schematic of (a) spherical indentation and (b) conical indentation. The spherical indenter has a diameter of $d = 7.94$\,mm, and the conical indenter has a base angle of $ 30^{\circ}$. Both specimens have dimensions of $H = 12.9$\,mm and $R_0 = 14.4$\,mm. Indenter force $P$ versus indenter displacement $\delta$ for (c) spherical indentation and (d) conical indentation for experimental data (solid black lines with shaded error regions) and simulations using material parameter sets from VFM-GA  (dashed orange lines) and manual fitting (dashed blue lines) for all three densities of Poron XRD foam.}
    \label{fig:1st_indent}
\end{figure}

\begin{figure}[!t]
    \centering
    \includegraphics[width=1\textwidth]{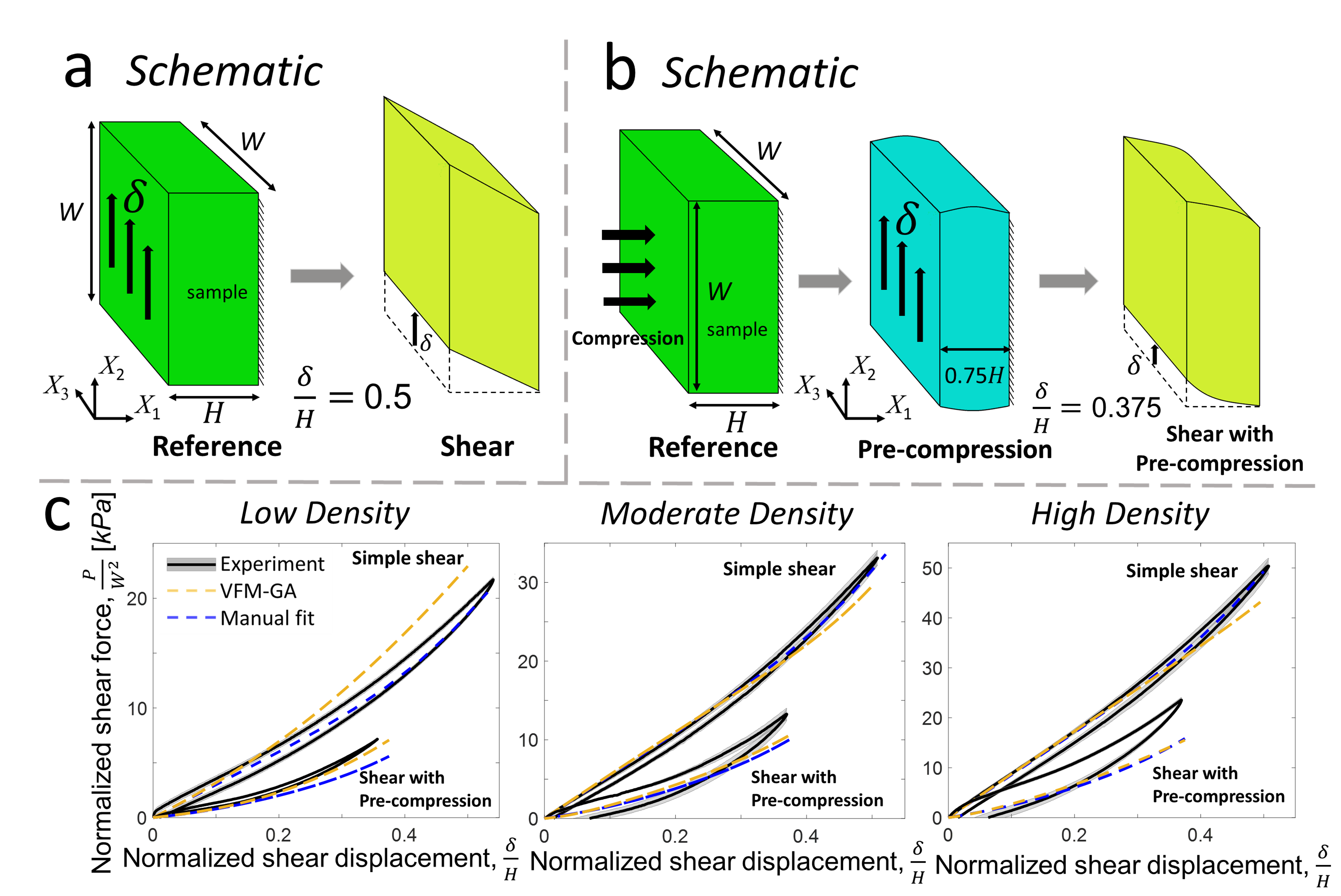}
    \caption{Validation test for the 1st-func: Simple-shear-like deformation without and with pre-compression. Schematic of (a) shear and (b) shear with pre-compression, both using a specimen with dimensions of $H = 12.9$\,mm and $W = 30$\,mm. In (a), a normalized shear displacement of $\delta/H = 0.5$ is applied on the left surface of the specimen. In (b), the specimen is first pre-compressed until the original width $H$ is decreased to $0.75H$. A normalized shear displacement of $\delta/H = 0.375$ is then applied to the pre-compressed specimen. In (c), the corresponding histories of normalized shear force $P/W^2$ versus normalized shear displacement $\delta/H$ are plotted for experimental data (solid black lines with shaded error regions) and simulations using material parameter sets from VFM-GA (dashed orange lines) and manual fitting (dashed blue lines) for Poron XRD foams of all three densities.}
    \label{fig:1st_shear}
\end{figure}

\begin{figure}[!t]
    \centering
    \includegraphics[width=0.7\textwidth]{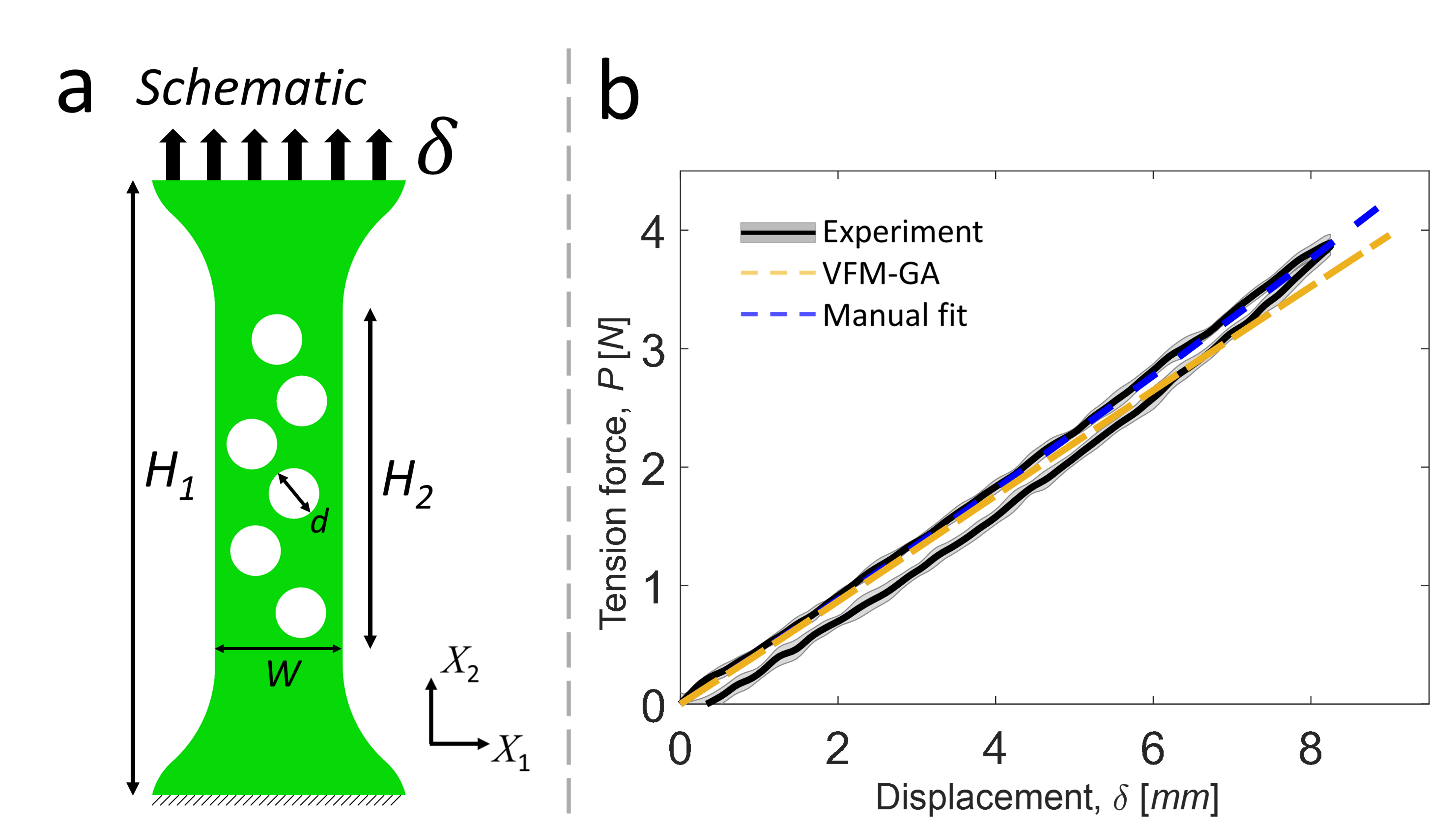}
    \caption{Validation test for the 1st-func: Tension of a specimen with circular holes. (a) Schematic of a dogbone specimen with dimensions of  $H_1 = 60.8$\,mm, $H_2 = 33.6$\,mm, and $W = 12.9$\,mm. Six circular holes, each with a diameter of $d = 5$\,mm, are cut through the dogbone specimen. The displacement $\delta$ is applied uniformly on the top surface. (b) Tension force $P$ versus displacement $\delta$ for experimental data (solid black lines with shaded error regions) and simulations using material parameter sets from VFM-GA (dashed orange lines) and manual fitting (dashed blue lines) for the high-density Poron XRD foam.}
    \label{fig:1st_inhomo}
\end{figure}

Next, it is necessary to evaluate the ability of the VFM-GA material parameter sets to predict the response of foam specimens undergoing inhomogeneous deformation to confirm that the VFM-GA parameter sets provide the same or better predictive capability as the manually fitted parameter sets of \citet{landauer2019experimental}. Three inhomogeneous validation tests are conducted: (1) spherical and conical indentation (Fig.~\ref{fig:1st_indent}), (2) simple-shear-like deformation both without and with pre-compression (Fig.~\ref{fig:1st_shear}), and (3) tension of a specimen with circular holes (Fig.~\ref{fig:1st_inhomo}). The experimental data and simulation results using the manually fit parameters are taken from \citet{landauer2019experimental}. The specific methodologies of the experimental setups and the finite-element simulations in \citet{abaqus} are documented in that paper and are, therefore, not described in detail here. In this study, Abaqus simulations of the same experiments are performed with the VFM-GA material parameter sets, and the simulated results are compared to the experimental data and the simulation results using the manually fit parameters to evaluate the performance of the VFM-GA parameters. 

\paragraph{Spherical and conical indentation} Figures \ref{fig:1st_indent}(a) and (b) illustrate schematics of the spherical and conical indentation validation tests, respectively. In both cases, the height of the specimen is $H=12.9$\,mm, and an axisymmetric geometry with a specimen radius of $R_0=14.4$\,mm is used in the finite-element simulations. To model the fixed-base boundary
condition, all displacement components are prescribed to be zero on the bottom face of the mesh, and the lateral surface
of the mesh is traction-free. The displacement $\delta$ along the $X_1$-direction is applied to an axisymmetric spherical indenter of diameter $d=7.94$\,mm or a conical indenter with a base angle of $30^{\circ}$ to indent the top surface of the specimen. The indenter force versus displacement responses in experiments and simulations are then compared for all three densities of Poron foams in Figs.~\ref{fig:1st_indent}(c) and (d) for spherical and conical indentation, respectively. Model predictions using the VFM-GA parameters and the manually fit parameters are shown using dashed orange and blue lines, respectively, and the experimental data is shown with solid black lines with shaded error regions. The VFM-GA material parameter sets perform similarly to the manually fit parameters, capturing the experimental load paths with a comparable amount of error for each density of Poron and for both shapes of indenters. 

\paragraph{Simple-shear-like deformation without and with pre-compression}
In the simple-shear-like validation tests, we start with a cuboid sample of thickness $H$ along the $X_1$-direction and height and depth $W$ along the $X_2$- and $X_3$-directions. The specimen has all degrees of freedom on its right surface prescribed to be zero. As illustrated in Fig.~\ref{fig:1st_shear}(a), a displacement $\delta$ is applied uniformly on the left surface along the $X_2$-direction, while maintaining the displacements along the $X_1$- and $X_3$-directions to be zero. The lateral surfaces are traction-free. The result is simple-shear-like deformation in the specimen, which is inhomogeneous due to edge effects near the traction-free boundaries. Pre-compression may also be applied prior to shear, as illustrated in Fig.~\ref{fig:1st_shear}(c), in which a displacement of $0.25H$ is first applied uniformly on the left surface of the specimen along the $X_1$-direction. Then, the pre-compression is held fixed while the displacement $\delta$ is applied along the $X_2$-direction, resulting in a state of deformation that couples shear and compression. In both types of tests, the normalized shear force ${P}/{W^2}$ is recorded as a function of the normalized shear displacement ${\delta}/{H}$ along the load path. Figure~\ref{fig:1st_shear}(c) illustrates comparisons of the $P$ versus $\delta$ relations between experiments and simulations using both VFM-GA and manually fit material parameter sets for each density of Poron XRD foam. In these two shear-related validation tests, the VFM-GA material parameter sets exhibit performance comparable to the manually fit parameter sets, performing slightly better for the cases with pre-compression and slightly worse for the cases without pre-compression.

\paragraph{Tension of a specimen with circular holes}
A schematic of a dogbone specimen (ASTM D3574 standard geometry) with six circular holes (each with a diameter of $d=5$\,mm) is illustrated in Fig.~\ref{fig:1st_inhomo}(a). The specimen is modeled in three dimensions with an out-of-plane thickness of 12.9\,mm. All degrees of freedom on the bottom surface of the specimen are fixed, and a displacement $\delta$ along the $X_1$-direction is uniformly applied on the top surface, while the displacements along the $X_2$- and $X_3$-directions are fixed. All other surfaces are traction free. The result is  inhomogeneous deformation in the tension specimen. The tensile force $P$ as a function of the displacement $\delta$ is recorded along the load path. In this validation test, we focus on the high-density Poron XRD foam, and Fig.~\ref{fig:1st_inhomo}(b) shows a comparison between $P$ versus $\delta$ relations from experiments and simulations using both VFM-GA and manually fit parameter sets. Again, the response predicted by the VFM-GA material parameter set exhibits comparable performance to the manually calibrated set. 

\subsection{Application of the 2nd-func}\label{sec:2nd}
Next, we consider two applications of the 2nd-func of the VFM-GA framework, focusing on homogeneous and inhomogeneous deformation experiments, respectively. The first application uses full-field DIC displacement fields and synchronized load-cell data from homogeneous simple compression and tension experiments. The second application uses data from a wedge indentation experiment, in which the deformation fields are inhomogeneous. All experimental data used in this section is new experimental data generated for the application of the 2nd-func using the experimental setup described in \citet{landauer2019experimental}. We focus on the high-density Poron XRD foam, and full-field displacement data for all experiments is generated using the STAQ-DIC algorithm of \citet{Yang2022}. 

\subsubsection{Application of the 2nd-func to homogeneous simple compression and tension}\label{sec:2nd_homo}

\begin{figure}[!t]
    \centering
    \includegraphics[width=0.9\textwidth]{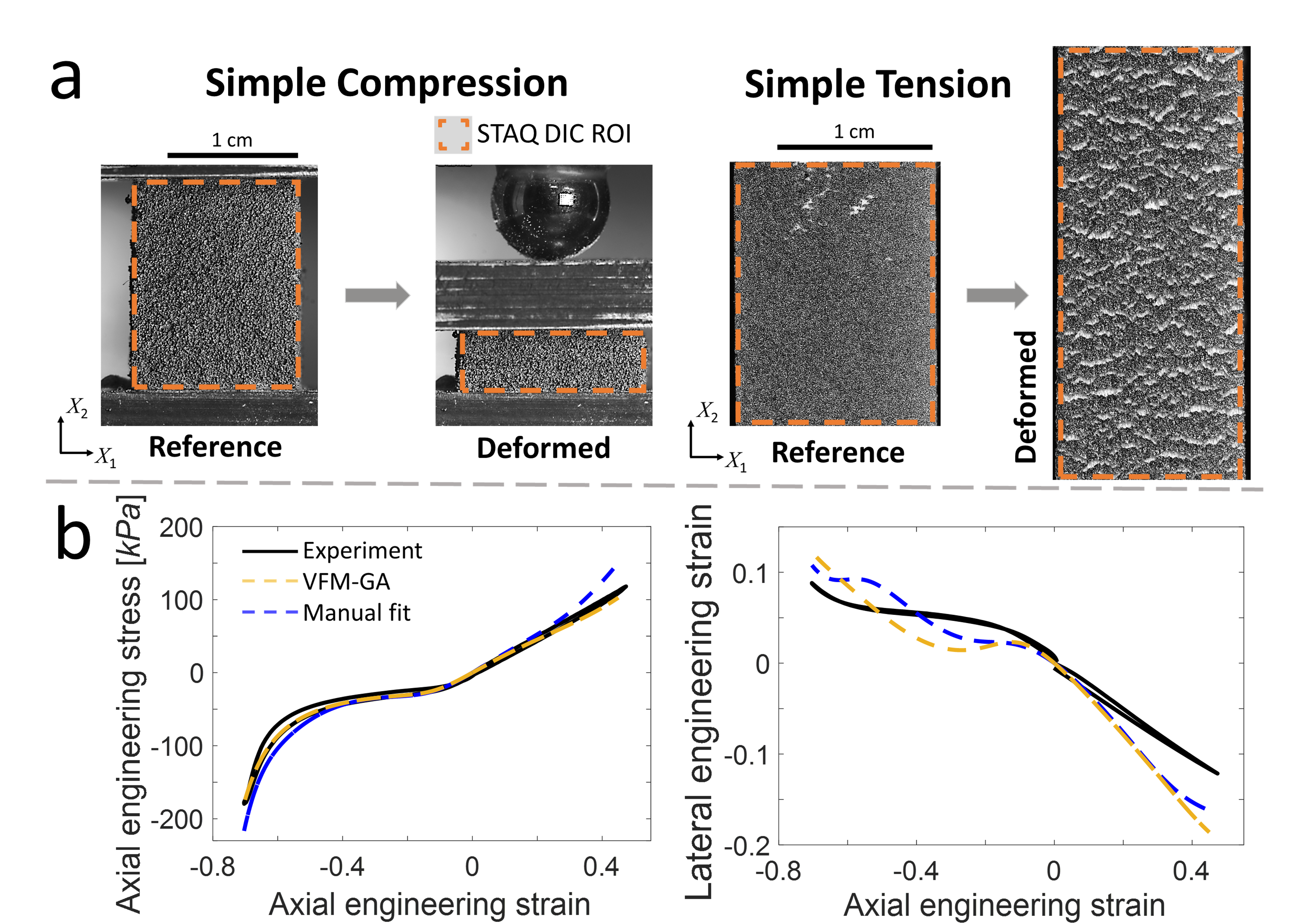}
    \caption{Results for the 2nd-func in simple compression and tension for high-density Poron XRD foam. (a) Images of the simple compression (left) and simple tension (right) specimens in both the reference and deformed states. The dashed orange contours indicate the STAQ-DIC ROI which adapts as the specimen deforms. In simple compression, the specimen was compressed to an engineering strain of $-0.7$ at an engineering strain rate magnitude of $2\times 10^{-4}$\,s$^{-1}$. In simple tension, the specimen was deformed to engineering strain of $0.45$ at the same engineering strain rate magnitude. (b) Comparisons of the engineering stress-strain curves (left) and engineering lateral-axial strain curves (right) between experimental data (solid black lines) and constitutive model predictions using the VFM-GA material parameters of Table~\ref{tab:fit_params_2nd} (dashed orange lines) and the manually fit parameters of \citet{landauer2019experimental} (dashed blue lines).}
    \label{fig:2nd_homo}
\end{figure} 

\paragraph{VFM inputs} The input for this application is data from the homogeneous simple compression and tension experiments illustrated in Fig.~\ref{fig:2nd_homo}(a). The simple compression experiment uses a cuboid specimen made of high-density Poron XRD foam with a width along the $X_1$-direction of $10.4$\,mm, a height along the $X_2$-direction of $12.8$\,mm, and an out-of-plane depth of $10.4$\,mm. The bottom surface of the specimen is placed on a fixed platen, and compression under displacement control with an engineering strain rate magnitude of $2\times 10^{-4}$\,s$^{-1}$ is applied to the top surface of the specimen to an engineering strain of $-0.7$. The top and bottom surfaces of the specimen are lubricated as described in \citet{landauer2019experimental} to minimize barreling. The flexible DIC ROI covers the entire front surface of the specimen throughout the load path, and DIC images for 17 load steps between engineering strains of $-0.08$ and $-0.7$, along with their synchronized load cell data, are sampled evenly based on time. The simple tension experiment uses a dogbone specimen of high-density Poron XRD foam. The initial, rectangular DIC ROI has a width of $12.2$\,mm and a height of $13.4$\,mm, and the specimen has an out-of-plane depth of $13.4$\,mm. The top and bottom surfaces of the dogbone specimen, which are not pictured in the figure, are clamped. Tension under displacement control is applied on the top surface at an engineering strain rate of $2\times 10^{-4}$\,s$^{-1}$ to an engineering strain of $0.45$. The flexible DIC ROI covers the front rectangular region of the specimen throughout the load path, and DIC images for 13 load steps between engineering strains of $0.05$ and $0.45$ are extracted based on even time sampling. 

\paragraph{GA inputs} The prescribed ranges for the material parameters in this application of the 2nd-func are slightly different from the ranges used in the application of the 1st-func in Section~\ref{sec:1st}. Specifically, the ground-state shear modulus $G_0$ and bulk modulus $B$ are not fixed because the input to the 2nd-func does not include stress-strain and lateral-axial strain curves. In this 2nd-func application, we prescribe the ranges for $G_0$ and $B$ to be $\pm 20\%$ of the values used in the 1st-func application in Section~\ref{sec:1st}. The prescribed ranges for the other 12 parameters remain the same as in Section~\ref{sec:1st}. The prescribed material parameter ranges in this application are documented table in Tables~\ref{tab:prescribed_12} and \ref{tab:prescribed_gb} in Appendix~\ref{app:mat para}. The weight in the objective function, the number of load steps, and the stability filter parameters are documented in Table \ref{tab:VFM_parameters} of Appendix~\ref{app:code para}, and the GA hyperparameters are documented in Table \ref{tab:GA_hyperparameters} of Appendix~\ref{app:hyper para}. The plane-stress solver of Section \ref{sec:stress_field} is used in this application of the 2nd-func. 

\begin{table}[!t]
\setlength{\tabcolsep}{3.95pt} 
\centering
 \begin{tabular}{cccccccccccccccc}
 \toprule
 Density & Fit & $G_0$ [kPa] & $B$ [kPa] & $J_{\rm min}$ & $C_1$ & $K_1^0$ & $\Delta_{\rm K}$ & $X_1^{\prime}$ & $X_2^{\prime}$ & $C_0$ & $p$ & $q$ & $r$ & $C_2$ & $C_3$ \\ 
 \noalign{\smallskip}\hline\noalign{\smallskip}
 \multirow{2}{*}{High}& Manual & 102 & 193.8 & 0.19 & 1.9 & -0.21 & 0.2 & 3.7 & 0.22 & 0.1 & 4 & 5 & 2 & 9 & 0.026 \\
 & VFM-GA & 92.6 & 223.3 & 0.21 & 0.84 & -0.133 & 0.307 & 4.13 & 0.208 & 0.05 & 4.8 & 4.8 & 1 & 17  & 0.167 \\
 \hline
 \end{tabular}
\caption {Comparison of material parameters for high density Poron XRD foam determined using the 2nd-func of the VFM-GA framework applied to homogeneous simple compression and tension versus manual fitting.} 
\label{tab:fit_params_2nd}
\end{table}

\paragraph{VFM-GA results}
Table~\ref{tab:fit_params_2nd} shows the  material parameters obtained from applying the 2nd-func of the VFM-GA framework to homogeneous simple compression and tension and compares them with the manually fitted parameters from \citet{landauer2019experimental}. Using the VFM-GA 2nd-func material parameter set of Table~\ref{tab:fit_params_2nd}, we calculate engineering stress-strain and engineering lateral-axial strain curves, which are compared with experimental data in Fig.~\ref{fig:2nd_homo}(b) for the high-density Poron XRD foam. The experimental data is shown using solid black lines, and calculations using the VFM-GA and manually fit parameters are shown using dashed orange and blue lines, respectively. Similar to the application of the 1st-func in Section \ref{sec:1st_2}, the calibrated material parameter set from this application of the 2nd-func outperforms the manually fit parameters in simple compression and tension. Several of the validation tests of Section~\ref{sec:1st_3} were repeated using the VFM-GA material parameter set of Table~\ref{tab:fit_params_2nd}, and the predictive capability of the parameter set is similar to that demonstrated in Figs.~\ref{fig:1st_indent}, \ref{fig:1st_shear}, and \ref{fig:1st_inhomo}. Therefore, these results are omitted here for brevity. 

\subsubsection{Application of the 2nd-func to wedge indentation}\label{sec:2nd_inhomo}

\begin{figure}[!t]
    \centering
    \includegraphics[width=1.0\textwidth]{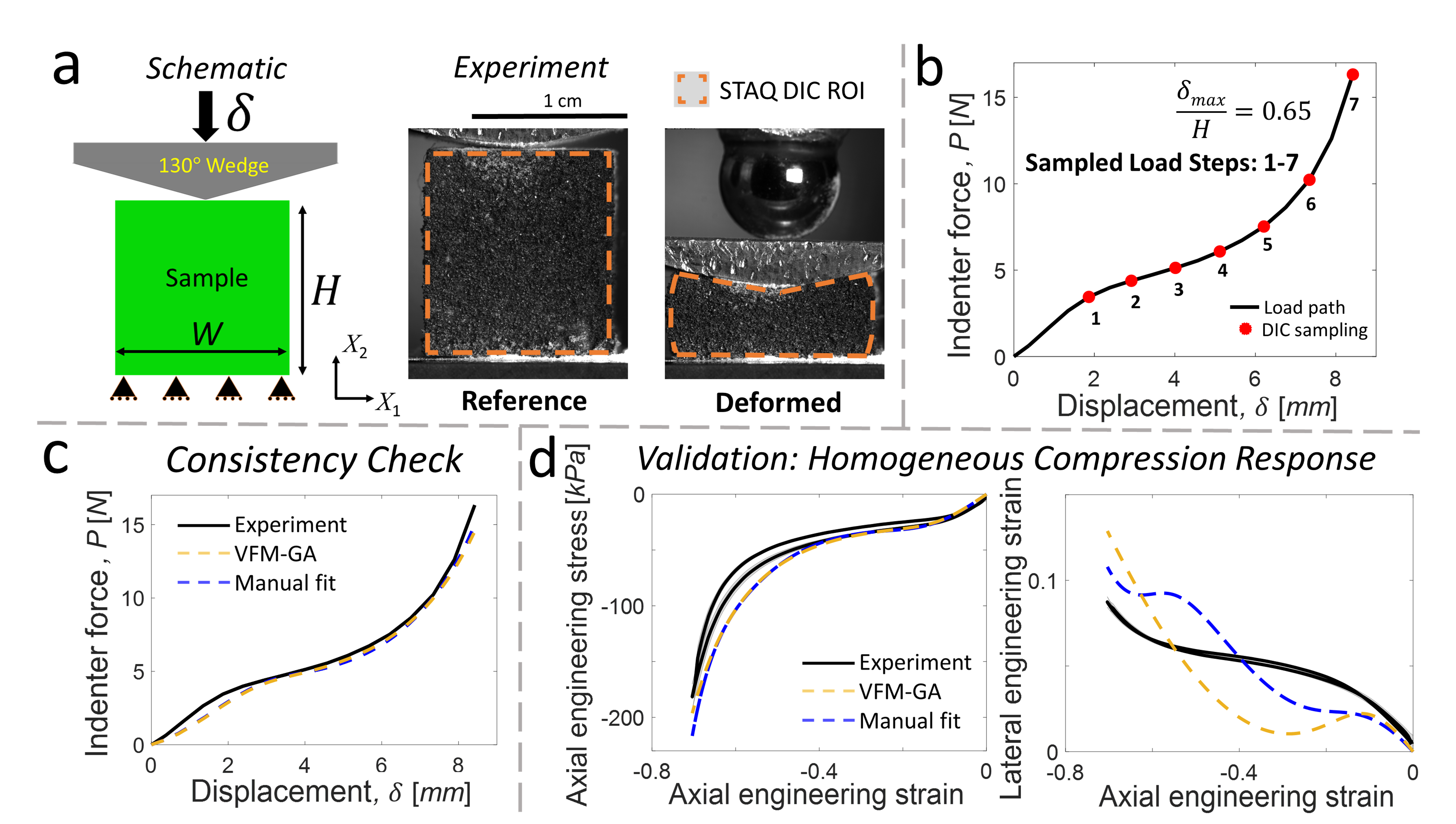}
    \caption{Results for the 2nd-func applied to wedge indentation of the high-density Poron XRD foam. (a) Schematic of the wedge indentation experiment with images of the reference and deformed configurations. The specimen has width $W = 12.3$\,mm, height $H = 12.9$\,mm, and out-of-plane depth $12.2$\,mm, and the wedge indenter has an included angle of $130^\circ$. The dashed orange contours indicate the STAQ-DIC ROI. (b) Experimental indenter force $P$ versus wedge tip displacement $\delta$. The seven sampled load steps serving as the input for the VFM-GA framework are marked using red circles along the load path. (c) Consistency check, in which the experimental load path (solid black line) is compared against constitutive model predictions using the VFM-GA material parameters of Table~\ref{tab:fit_params_2nd_in} (dashed orange line). (d) Validation test, in which engineering stress-strain and lateral-axial strain curves for compression are calculated using the VFM-GA material parameters of Table~\ref{tab:fit_params_2nd_in} (dashed orange lines) and compared against the experimental data (solid black lines). For reference, The VFM-GA results are also compared against the results using the manually fit parameters of \citet{landauer2019experimental} (dashed blue lines) in (c) and (d).}
    \label{fig:2nd_inhomo}
\end{figure}

\paragraph{VFM inputs} Next, we use data from a wedge indentation experiment to test the capability of the 2nd-func of the VFM-GA framework when applied to inhomogeneous deformation fields. A schematic of the experiment is illustrated in Fig.~\ref{fig:2nd_inhomo}(a). The experiment involves a cuboid specimen of high-density Poron XRD foam with width $W = 12.3$\,mm, height $H = 12.9$\,mm, and out-of-plane depth $12.2$\,mm. The top and bottom surfaces of the specimen are lubricated, and the specimen is then placed on fixed, flat platen. An aluminum wedge with an included angle of $130^{\circ}$ and a uniform out-of-plane shape with depth greater than that of the specimen serves as the indenter. The vertical centerline of specimen is aligned with the tip of the wedge indenter creating a symmetric geometry. The four lateral surfaces of the specimen are traction-free. The displacement $\delta$ along the negative $X_2$-direction in Fig.~\ref{fig:2nd_inhomo}(a) is applied to the wedge indenter at a speed of $2\times10^{-3}$\,mm/s to a maximum displacement of $\delta/{H} = 0.65$. The images in Fig.~\ref{fig:2nd_inhomo}(a) illustrate the reference configuration and the final deformed configuration with the dashed orange lines indicating the STAQ-DIC ROI. This experimental design results in inhomogeneous deformation fields that are approximately the same for any $X_1$-$X_2$ cross-section of the specimen due to the nearly frictionless interaction between the specimen and the indenter and bottom platen. The experimental load path, i.e., the indenter force $P$ versus the wedge tip displacement $\delta$, is plotted in Fig.~\ref{fig:2nd_inhomo}(b) using a solid black line. The red dots along the load path are the load steps that are selected to be used in the 2nd-func of the VFM-GA framework. Because inhomogeneous deformation fields contain richer information regarding the deformation states compared to homogeneous fields, we only sample 7 load steps marked by the numbers 1 to 7 in Fig.~\ref{fig:2nd_inhomo}(b). These 7 load steps of inhomogeneous DIC displacement fields with synchronized indenter force data serve as the experimental input for the second application of the 2nd-func of the VFM-GA framework, focusing on the compression response only. 

\paragraph{GA inputs} The prescribed ranges for the material parameters in this application of the 2nd-func remain the same as the ranges used in Section \ref{sec:2nd_homo} and are documented in Tables~\ref{tab:prescribed_12} and \ref{tab:prescribed_gb}. The other VFM-related parameters and GA hyperparameters are documented in Tables \ref{tab:VFM_parameters} and \ref{tab:GA_hyperparameters}, respectively. Besides the number of load steps, the weight in the objective function $\alpha$, and the use of the monotonicity filters, the other parameters are unchanged from those used in the first application of the 2nd-func in Section \ref{sec:2nd_homo}. 

\begin{table}[!t]
\setlength{\tabcolsep}{3.95pt} 
\centering
 \begin{tabular}{cccccccccccccccc}
 \toprule
 Density & Fit & $G_0$ [kPa] & $B$ [kPa] & $J_{\rm min}$ & $C_1$  & $K_1^0$ & $\Delta_{\rm K}$ & $X_1^{\prime}$ & $X_2^{\prime}$ & $C_0$ & $p$ & $q$ & $r$ & $C_2$ & $C_3$ \\ 
 \noalign{\smallskip}\hline\noalign{\smallskip}
 \multirow{2}{*}{High}& Manual & 102 & 193.8 & 0.19 & 1.9 & -0.21 & 0.2 & 3.7 & 0.22 & 0.1 & 4 & 5 & 2 & 9 & 0.026 \\
 & VFM-GA & 95.3 & 233.8 & 0.1 & 0.05 & 0 & 0.377 & 6.27 & 0.076 & 0.05 & 3.6 & 5.2 & 1 & 18.3 & 0.367 \\
 \hline
 \end{tabular}
\caption{Comparison of material parameters for high density Poron XRD foam determined using the 2nd-func of the VFM-GA framework applied to wedge indentation versus manual fitting.} 
\label{tab:fit_params_2nd_in}
\end{table}

\paragraph{VFM-GA results} Table~\ref{tab:fit_params_2nd_in} compares the material parameters obtained from applying the 2nd-func of the VFM-GA framework to wedge indentation with the manually fitted parameters. Before testing the parameter set on independent experimental data, we first perform a `consistency check', in which we verify that this material parameter set can adequately capture the full load path in wedge indentation. To this end, three-dimensional Abaqus simulations of the wedge indentation experiment are performed using both sets of material parameters in Table~\ref{tab:fit_params_2nd_in}. Due to the symmetric nature of the problem, a quarter of the sample (half along the width and half along the depth) is modeled in Abaqus using 1024 C3D8H elements. Symmetry boundary conditions are applied along the symmetry planes, and the remaining two lateral surfaces are subjected to traction-free boundary conditions. The wedge indenter and bottom platen are idealized as rigid surfaces, and the displacement $\delta$ is applied to the wedge indenter, while the bottom platen is held fixed. The friction coefficient between the specimen and each surface is taken to be 0.15, consistent with the lubricated condition. Figure \ref{fig:2nd_inhomo}(c) compares the load paths from the experiment (solid black line) and simulations using the VFM-GA (dashed orange line) and manually fit (dashed blue line) material parameter sets. The simulated load paths for the VFM-GA and manually fit parameter sets nearly overlap with each other, both capturing the experimental load path reasonably well. However, one may expect the VFM-GA parameter set to outperform the manually fit parameter set since it was calibrated using data from the wedge indentation experiment, while the manually fit set was calibrated to simple compression and tension data. This observation may potentially be explained by the non-uniform out-of-plane deformation around the wedge tip observed in both experiment and simulations. Such non-uniform out-of-plane deformation can slightly distort the DIC field measurements, leading to an imperfect VFM-GA calibration. 

\paragraph{Validation test} Finally, since experimental data from homogeneous compression was not used to determine the VFM-GA parameter set of Table~\ref{tab:fit_params_2nd_in}, engineering stress-strain and lateral-axial strain curves for simple compression of the high-density Poron XRD foam are calculated using the VFM-GA parameters as a validation test. The results are shown in Fig.~\ref{fig:2nd_inhomo}(d) with the experimental data shown using solid black lines and model predictions using the VFM-GA and manually fit parameters shown using dashed orange and blue lines, respectively. The VFM-GA and manually fit parameter sets perform similarly in capturing the stress-strain response. For the lateral-axial strain response, the performance of the VFM-GA parameter set of Table~\ref{tab:fit_params_2nd_in} is reasonable but not as accurate as the manual fit. This is because the manually fit parameters were calibrated directly to the simple compression data, while this VFM-GA parameter set was calibrated to the independent wedge indentation data. Moreover, this VFM-GA parameter set performs comparably to the VFM-GA material parameter set of Table~\ref{tab:fit_params_2nd} determined in Section~\ref{sec:2nd_homo} (although slightly less accurately for the stress-strain response), while using fewer input load steps (seven) than the application of Section~\ref{sec:2nd_homo} (17 homogeneous deformation fields in compression). The application of the 2nd-func of the VFM-GA framework in this section demonstrates its ability to determine material parameters based on experimental data for inhomogeneous deformation fields and to do so efficiently with fewer load steps when using inhomogeneous deformation fields as input. 

\section{Noise study and the effect of the weight parameter $\alpha$}\label{sec:noise}
To understand how noise in the experimental DIC displacement fields affects the performance of the VFM-GA framework, in this section, we carry out a noise study using synthetically generated displacement fields corresponding to simple compression/tension. The ideal displacement fields are perturbed by different noise levels, and the 2nd-func of the VFM-GA is then applied to each data set. The resulting material parameter sets for each noise level are evaluated based on their ability to capture corresponding ideal engineering stress-strain and lateral-axial strain curves. Moreover, we examine the effect of the weight parameter $\alpha$ in \eqref{obj} and its interplay with the noise level by carrying out the noise study for three different values of $\alpha$. 

\subsection{Noise study inputs}\label{sec:noise_input} 
\begin{figure}[!t]
\centering
    \includegraphics[width=0.58\textwidth]{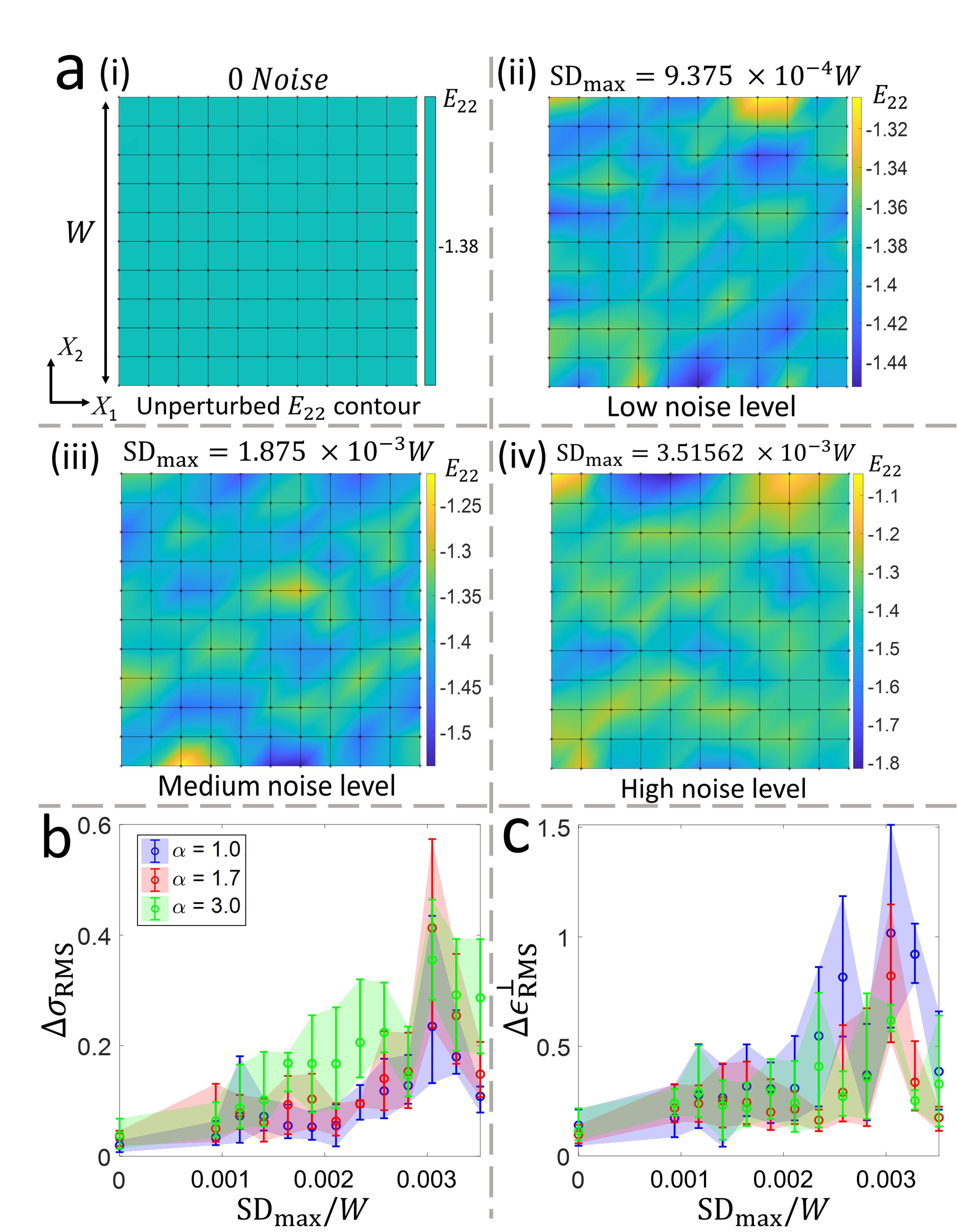}
    \caption{Results for the noise study. (a) Local contours of the logarithmic strain component $E_{22}$ corresponding to input displacement fields for a far-field axial logarithmic strain of $-1.38$ with (i) no noise, (ii) a low noise level (${\rm SD}_{\rm max} = 9.375\times 10^{-4}W$), (iii) a medium noise level (${\rm SD}_{\rm max} = 1.875\times 10^{-3}W$), and (iv) a high noise level (${\rm SD}_{\rm max} = 3.51562\times 10^{-3}W$). (b) $\Delta \sigma_{\rm RMS}$ and (c) $\Delta \epsilon_{\rm RMS}^{\perp}$ in simple compression/tension versus the normalized noise level ${\rm SD}_{\rm max}/W$ for material parameter sets determined using the 2nd-func of the VFM-GA framework. In (b) and (c), 13 different noise levels and three different values of the weight parameter $\alpha$ are considered. Blue, red, and green markers denote results for $\alpha = 1.0, 1.7$, and $3.0$, respectively. The markers represent the mean values of $\Delta \sigma_{\rm RMS}$ and $\Delta \epsilon_{\rm RMS}^{\perp}$ over the top five material parameter sets for each combination of ${\rm SD}_{\rm max}$ and $\alpha$, and the error bars indicate the maximum and minimum values among the same five sets. Colored patches that connect the top and bottom ends of consecutive error bars for the same value of $\alpha$ are included as a guide for the eye.}
    \label{fig:noise_study}
\end{figure}

To construct the synthetic datasets employed in the noise study, we use the VFM-GA material parameter set for the medium-density foam, outlined in Table \ref{tab:1st_paras}. This baseline material parameter set is used to compute the corresponding axial stress and lateral strain histories as a function of the axial logarithmic strain $E_{22}$ in simple compression/tension. We then utilize a square domain with side length $W$ to represent the reference configuration and discretize the domain with 100 square elements, as shown in Fig.~\ref{fig:noise_study}(a),(i). For a given load step, the corresponding axial and lateral strains are used to calculate the ideal nodal displacements under the assumption that deformation is homogeneous, and the reaction forces along the top and bottom boundaries are calculated using the axial stress for that load step. We generate 16 load steps in simple compression with $E_{22}$ ranging from -0.01 to -1.38 and 9 load steps in simple tension with $E_{22}$ ranging from 0.01 to 0.34. These 25 load steps, encompassing displacement fields and corresponding reaction forces, alongside the undeformed mesh, form the ideal, zero-noise dataset that is compatible with the 2nd-func of the VFM-GA framework.  

Next, to create datasets with varying noise levels, we additively perturb each nodal displacement component in each load step. The perturbation for each nodal displacement is drawn from a Gaussian distribution centered about zero and with standard deviation ${\rm SD}$. To realistically approximate noise in DIC measurements, we base our approach on the results of \citet{landauer2018} (specifically Fig.~5 in that work), which demonstrated that the noise in the DIC-computed displacement fields in simple compression increases with increasing axial strain magnitude. To account for this effect, we take the standard deviation ${\rm SD}$ of the Gaussian distribution to be a linear function of the axial logarithmic strain magnitude:
\begin{equation}\label{eqn:SD of noise study}
{\rm SD}(E_{22}) = \dfrac{{\rm SD}_{\rm max}-{\rm SD}_{\rm min}}{E_{\rm max}-E_{\rm min}}(|E_{22}|-E_{\rm min})+{\rm SD}_{\rm min}\quad\text{for}\quad |E_{22}|>E_{\rm min},
\end{equation}
where ${\rm SD}_{\rm max}$ and ${\rm SD}_{\rm min}$ are the standard deviations at $E_{22}$ magnitudes of $E_{\rm max}$ and $E_{\rm min}$, respectively, which we take to be $E_{\rm max}=1.38$ and $E_{\rm min}=0.01$. Utilizing the Fast Interative DIC results in Fig.~5 of \citet{landauer2018}, we non-dimensionalize their displacement errors at the same axial strain magnitudes and convert them to ${\rm SD}_{\rm max}=9.375\times 10^{-4}W$ and ${\rm SD}_{\rm min} = 1.95\times 10^{-6}W$. This establishes the baseline noise level. We also consider 11 additional noise-level cases by maintaining ${\rm SD}_{\rm min}$ constant and increasing ${\rm SD}_{\rm max}$ up to ${\rm SD}_{\rm max} = 3.51562\times 10^{-3} W$ to account for higher noise levels that could potentially be encountered in experiments. Figure~\ref{fig:noise_study}(a),(i) depicts the local contours of $E_{22}$ on the undeformed geometry for a far-field axial strain of $E_{22} = -1.38$ without any noise, illustrating the state of homogeneous deformation. Figures~\ref{fig:noise_study}(a),(ii)-(iv) illustrate the local contours of $E_{22}$ for the same far-field axial strain of $E_{22} = -1.38$ for noise levels of  ${\rm SD}_{\rm max} = 9.375\times 10^{-4}W$, ${\rm SD}_{\rm max} = 1.875\times 10^{-3} W$, and ${\rm SD}_{\rm max} = 3.51562\times 10^{-3}W$. We classify these three cases as low, medium, and high noise levels, respectively. In total, 13 sets of displacements fields (one without noise and 12 with different noise levels) serve as the inputs for the 2nd-func of the VFM-GA framework. 

Moreover, to study the effect of the weight parameter $\alpha$ in the objective function \eqref{obj} and its interplay with the noise level, the 2nd-func of the VFM-GA framework is applied three times with $\alpha = 1.0, 1.7$, and $3.0$ for each set of displacement fields, while not changing any other input parameters (39 cases in total). For each case, the five calibrated material parameter sets that correspond to the lowest five objective function values in the final generation are collected for analysis (five parameters sets for each of the 39 cases giving 195 material parameter sets in total). The prescribed ranges for the material parameters in the noise study are documented in Tables \ref{tab:prescribed_12} and \ref{tab:prescribed_gb}, and the VFM-related parameters and GA hyperparameters are documented in Tables \ref{tab:VFM_parameters} and \ref{tab:GA_hyperparameters}, respectively. 

\subsection{Results}\label{sec:noise_result} 
To quantitatively evaluate each calibrated material parameter set, we compute its axial engineering stress and lateral engineering strain histories in simple compression/tension and compare with the ideal histories calculated using the baseline material parameter set. Denote the axial engineering strain as $\epsilon$, the ideal and VFM-GA axial engineering stress histories as $\sigma_{\rm ideal}(\epsilon)$ and $\sigma_{\rm GA}(\epsilon)$, and the ideal and VFM-GA lateral engineering strain histories as $\epsilon_{\rm ideal}^\perp(\epsilon)$ and $\epsilon_{\rm GA}^\perp(\epsilon)$. For each calibrated material parameter set, we calculate the normalized root-mean-square difference between its axial engineering stress and lateral engineering strain histories and the ideal histories, defined as
\begin{equation}\label{stress_error} 
\Delta \sigma_{\rm RMS} = \sqrt{\left. \int_{-0.7}^{0.4} \left(\sigma_{\rm GA}(\epsilon)-\sigma_{\rm ideal}(\epsilon)\right)^2 \, d\epsilon \middle/\int_{-0.7}^{0.4}(\sigma_{\rm ideal}\left(\epsilon)\right)^2\, d\epsilon \right. }
\end{equation}
and 
\begin{equation}\label{strain_error} 
\Delta \epsilon_{\rm RMS}^{\perp} = \sqrt{\left.\int_{-0.7}^{0.4}\left(\epsilon_{\rm GA}^{\perp}(\epsilon)-\epsilon_{\rm ideal}^{\perp}(\epsilon)\right)^2 \, d\epsilon \middle/ \int_{-0.7}^{0.4}\left(\epsilon_{\rm ideal}^{\perp}(\epsilon)\right)^2 \, d\epsilon\right.},
\end{equation}
where the axial engineering strain limits of -0.7 and 0.4 roughly correspond to the axial logarithmic strain limits of -1.38 and 0.34 over which the load steps were sampled. The integrals in \eqref{stress_error} and \eqref{strain_error} are evaluated numerically with an axial strain increment of 0.001. The non-dimensional quantities $\Delta \sigma_{\rm RMS}$ and $\Delta \epsilon_{\rm RMS}^{\perp}$ may then be used to quantify the performance of each calibrated material parameter set identified for analysis in the noise study. 

Figures~\ref{fig:noise_study}(b) and (c) illustrate the relationship between the displacement-field noise level ${\rm SD}_{\rm max}$ and the normalized differences $\Delta \sigma_{\rm RMS}$ and $\Delta \epsilon_{\rm RMS}^{\perp}$, respectively, for each of the three values of the weight parameter $\alpha$. For each combination of ${\rm SD}_{\rm max}$ and $\alpha$, the markers represent the mean values of $\Delta \sigma_{\rm RMS}$ and $\Delta \epsilon_{\rm RMS}^{\perp}$ over the top five calibrated material parameter sets for that combination, while the error bars indicate the maximum and minimum values among the same five sets. As a guide for the eye, colored patches are constructed by connecting the top and bottom ends of the error bars between consecutive noise levels for each $\alpha$ value with blue representing $\alpha = 1.0$, red representing $\alpha = 1.7$, and green representing $\alpha = 3.0$. Based on our experience, when $\Delta \sigma_{\rm RMS} \lesssim 0.125$ and $\Delta \epsilon_{\rm RMS}^{\perp} \lesssim 0.25$, a material parameter set adequately captures the ideal histories. All fifteen of the no-noise cases satisfy these criteria, while 11/15 of the low-noise cases, 6/15 of the medium noise cases, and 2/15 of the high-noise cases satisfy the criteria, indicating that the normalized differences tend to increase with the noise level, as expected. For ${\rm SD}_{\rm max}/W \lesssim 0.002$, the lower limits of the error bars, which are associated with the best-performing material parameter sets, remain below these criteria for all three values of $\alpha$, indicating that the VFM-GA framework can tolerate displacement fields with noise up to moderate levels. 

Regarding the effect of the weight parameter $\alpha$, for the no-noise and low-noise cases (${\rm SD}_{\rm max}/W \lesssim  0.001$), Figs.~\ref{fig:noise_study}(b) and (c) show overlap of the three colored patches, demonstrating that the effect of $\alpha$ on $\Delta \sigma_{\rm RMS}$ and $\Delta \epsilon_{\rm RMS}^{\perp}$ is small for low noise levels. However, as ${\rm SD}_{\rm max}$ increases to moderate and high noise levels, Fig.~\ref{fig:noise_study}(b) shows that $\alpha = 1.0$ performs better than $\alpha = 1.7$ and subsequently $\alpha = 3.0$ in terms of the normalized axial stress difference $\Delta \sigma_{\rm RMS}$. This trend is reversed for the normalized lateral strain difference $\Delta \epsilon_{\rm RMS}^{\perp}$, as shown in Fig.~\ref{fig:noise_study}(c). This is because $\Delta \epsilon_{\rm RMS}^{\perp}$ reflects how closely the balance of internal forces \eqref{balance of internal force 2} is satisfied, while $\Delta \sigma_{\rm RMS}$ reflects how closely the balance of external forces \eqref{balance of external force 3} is satisfied, so increasing the weight parameter $\alpha$ in the objective function \eqref{obj} biases the VFM-GA framework towards minimizing $\Delta \epsilon_{\rm RMS}^{\perp}$ over $\Delta \sigma_{\rm RMS}$ and vice versa as $\alpha$ is decreased. For low noise levels, this trade-off is not consequential, but for noisier input data,  the choice of $\alpha$ in VFM-GA framework becomes important. It is advisable to carefully choose $\alpha$ to balance this trade-off with the choice depending on the noisiness of the experimental data and the particular constitutive model under consideration. 

\section{Conclusion}\label{sec:conc}
In this paper, we introduced a VFM-GA calibration framework for isotropic hyperelastic constitutive models that integrates the Virtual Fields Method to define an objective function with a Genetic Algorithm as the optimization tool. The VFM-GA framework places no limitation on the functional form of the isotropic hyperelastic constitutive model and is designed to tackle the challenges of modeling nonlinear hyperelastic behavior, i.e., it can accommodate models with coupled dependencies on deformation invariants or a large number of material parameters. It also addresses the potential for material instability that can arise for certain hyperelastic constitutive models. The GA plays a pivotal role in tackling the optimization challenge with its robustness and adaptability in navigating a high-dimensional landscape. To broaden its applicability, the VFM-GA framework was implemented in two code packages with different functionalities, in which the first functionality uses experimental data from homogeneous simple compression/tension and the second functionality uses full-field DIC displacement fields and synchronized load cell data. Both VFM-GA functionalities were applied to real experimental input data for several densities of an open-cell elastomeric foam, using the nonlinear, large-deformation hyperelastic constitutive model of \citet{landauer2019experimental}.  The calibrated material parameter sets from the VFM-GA framework demonstrate improvements over manually fitted parameter sets from the literature across the tests, establishing the framework's proficiency in practical applications. 

Two pathways stand out for future work. The present version of the VFM-GA framework targets isotropic hyperelastic constitutive models, in which the stress and strain energy depend only on the state of deformation. Future work will focus on adapting the VFM-GA framework to history-dependent constitutive models, such as large-deformation viscoelastic constitutive models \citep{li2022large}, to widen the framework's applicability. In concert with the advancements in 3D displacement measurement in experimental mechanics \citep[e.g.,][]{estrada2020mr,wang2021inference}, a second avenue for future work is to extend the framework to 3D settings using experimental displacement fields determined using Digital Volume Correlation.

\section*{Acknowledgments}
This work was supported by funds from the US Office of Naval Research under PANTHER awards N00014-21-1-2044 and N00014-21-1-2855 through Dr. Timothy Bentley. 

\appendix
\renewcommand\thetable{A.\arabic{table}} 
\setcounter{figure}{0}
\setcounter{table}{0}
\section{Material parameter ranges, VFM parameters, and GA hyperparameters}
\subsection{Prescribed range of material parameters}\label{app:mat para}
The prescribed ranges of the material parameters for the constitutive model of Section~\ref{sec:model}, i.e., the sets $\boldsymbol{\theta}_{\rm min}$ and $\boldsymbol{\theta}_{\rm max}$, are given in Tables~\ref{tab:prescribed_12} and \ref{tab:prescribed_gb}. Table~\ref{tab:prescribed_12} lists the ranges of the twelve dimensionless material parameters which are consistent across Sections~\ref{sec:1st}, \ref{sec:2nd}, and \ref{sec:noise}. Table \ref{tab:prescribed_gb} lists the ranges of ground-state shear modulus $G_0$ and bulk modulus $B$ applied in each of the studies, which change across the cases. The specification of the ranges for each material parameter in a given constitutive model is an important step in applying the VFM-GA framework. Guidance may be taken from any bounds that must be placed on certain parameters. For example, in the constitutive model of Section~\ref{sec:model}, $0<J_{\rm min}<1$, so the specified range for $J_{\rm min}$ in Table~\ref{tab:prescribed_12} is consistent with this requirement. Guidance may also be taken from previous applications of the constitutive model, so that the specified ranges broadly span material parameter values reported in the literature, as was done in this work. When developing a new constitutive model, some iteration will likely be required to establish suitable ranges. For instance, if, after one trial execution of the VFM-GA framework, several of the calibrated parameters match their prescribed minimum or maximum values, one should adjust the ranges accordingly and perform another trial, repeating until the calibrated material parameters fall within their respective ranges. 

\begin{table}[!t]
\setlength{\tabcolsep}{4.pt} 
\centering
 \begin{tabular}{ccccccccccccc}
 \toprule
 Parameter range & $J_{\rm min}$ & $C_1$ & $K_1^0$ & $\Delta_{\rm K}$ & $X_1^{\prime}$ &  $X_2^{\prime}$ & $C_0$ & $p$ & $q$ & $r$ & $C_2$ & $C_3$ \\ 
 \noalign{\smallskip}\hline\noalign{\smallskip}
 $\boldsymbol{\theta}_{\rm min}$ & 0.1 & 0.5 & -0.5 & 0.05 & 2.0 & 0.01 & 0.01  & 2 & 2 & 1 & 5 & 0.001 \\
 $\boldsymbol{\theta}_{\rm max}$ & 0.3 & 6.0 & 0.0 & 0.4 & 10.0 & 1.0 & 6.0 & 8 & 8 & 6 & 25  & 0.5 \\
 \hline
 \end{tabular}
\caption {Prescribed ranges for the material parameters of the constitutive model of Section~\ref{sec:model}, except the ground-state moduli $G_0$ and $B$, used in the applications of the VFM-GA framework. The prescribed ranges listed in this table are used throughout this work.} 
\label{tab:prescribed_12}
\end{table}

\begin{table}[!t]
\setlength{\tabcolsep}{4.pt} 
\centering
 \begin{tabular}{ccccccc}
 \toprule
 \multirow{2}{*}{Parameter range} & 1st-func & 1st-func & 1st-func & 2nd-func & 2nd-func & \multirow{2}{*}{Noise study}\\
 & low density & medium density & high density & homogeneous & inhomogeneous & \\
 \noalign{\smallskip}\hline\noalign{\smallskip}
 min($G_0$) [kPa] & 34.5 & 65.2 & 102 & 82 & 82 & 52.16 \\
 max($G_0$) [kPa] & 34.5 & 65.2 & 102 & 122 & 122 & 78.24 \\
 \noalign{\smallskip}\hline\noalign{\smallskip}
 min($B$) [kPa] & 58.7 & 117.4 & 193.8 & 153.8 & 153.8 &  93.92 \\
 max($B$) [kPa] & 58.7 & 117.4 & 193.8 & 233.8 & 233.8 & 140.88 \\
 \hline
 \end{tabular}
\caption {Prescribed ranges for ground-state moduli $G_0$ and $B$ used in the applications of the 1st-func of the VFM-GA framework in Section~\ref{sec:1st} for low, medium and high density Poron XRD foams; the applications of the VFM-GA 2nd-func in Section~\ref{sec:2nd} for both homogeneous and inhomogeneous deformation input fields; and the noise study of Section~\ref{sec:noise}.} 
\label{tab:prescribed_gb}
\end{table}

\subsection{VFM-related parameters}\label{app:code para}
Table~\ref{tab:VFM_parameters} summarizes the VFM-related parameters used in the applications of the 1st-func of the VFM-GA framework in Section \ref{sec:1st}, in the applications of the VFM-GA 2nd-func in Section~\ref{sec:2nd}, and in the noise study of Section~\ref{sec:noise}. Each of the VFM-related parameters are discussed further below along with some guidance for their selection.

\begin{itemize}
    \item {$\alpha$}: The parameter $\alpha$ weighs the contribution of the internal force balance relative to the external force balance for a single load step in the objective function \eqref{obj}. As discussed in Section \ref{sec:noise_result}, in the context of simple compression/tension, the weight parameter $\alpha$ balances the performance of a calibrated material parameter set between the axial stress response and the lateral strain response. Based on our experience, when using the VFM-GA 1st-func, we recommend to start by choosing $2 \le \alpha \le 3$, and when using the VFM-GA 2nd-func, we suggest taking $1 \le \alpha \le 2$. Then, after applying the VFM-GA framework, one should inspect the resultant performance of the calibrated material parameter set and adjust $\alpha$ accordingly. The values summarized in Table~\ref{tab:VFM_parameters} were selected for the 1st-func and 2nd-func applications to obtain satisfactory calibration results.
    \item {$n_{\rm ten}$ and $n_{\rm comp}$}: The parameters $n_{\rm ten}$ and $n_{\rm comp}$ with $n_{\rm ten}+n_{\rm comp}=n_{\rm ls}$ are the number of load steps in tension and compression, respectively. As mentioned in Section~\ref{sec:1st_func_overview}, $n_{\rm ten}$ and $n_{\rm comp}$ should be selected carefully. This is because the contributions of each load step to the objective function \eqref{obj} are comparable, so selecting $n_{\rm comp}$ to be larger than $n_{\rm ten}$ biases the calibration process towards the compression response. Since the response of elastomeric foams in compression is more complex than the response in tension, such a bias is appropriate, and we take $n_{\rm comp}>n_{\rm ten}$ in all applications of the VFM-GA framework to simple compression/tension data, as summarized in Table~\ref{tab:VFM_parameters}. For the application of the VFM-GA 2nd-func to inhomogeneous deformation fields (Section~\ref{sec:2nd_inhomo}), we do not distinguish between $n_{\rm ten}$ and $n_{\rm comp}$ and simply take $n_{\rm ls}=7$. 
    \item{$(K_1,K_2)_{\rm c}$ and $(K_1,K_2)_{\rm s}$}: As discussed in Section~\ref{sec:loss_of_ellipticity}, $(K_1,K_2)_{\rm c}$ and $(K_1,K_2)_{\rm s}$ are the check points for the loss of ellipticity filter in compression $(K_3=-1)$ and shear $(K_3=0)$, respectively. To apply the loss of ellipticity filter, one calculates the material moduli at a check point identified by a set of invariants $\{K_1,K_2,K_3\}$ and tests the necessary and sufficient conditions for ellipticity of \citet{dacorogna2001}. The check points should be carefully chosen based on the potential for instability of a given constitutive model, and, for the constitutive model of Section~\ref{sec:model}, the values given in Table~\ref{tab:VFM_parameters} were selected to check the plateau regime in compression and the large deformation shear response. 
    \item{$C_{\rm mono}$ and $C_{\rm vol}$}: The parameters $C_{\rm mono}$ and $C_{\rm vol}$ are two Boolean variables that serve as flags for the filters of Section~\ref{sec:monotonic} that check the monotonicity of the axial engineering stress and volumetric strain histories, respectively, in simple compression/tension. If the VFM-GA framework is applied to input experimental data for homogeneous simple compression/tension using either the 1st-func or the 2nd-func, it is recommended to set $C_{\rm mono}=C_{\rm vol}=1$, turning both monotonicity filters on. Since these filters are only compatible with homogeneous simple compression/tension inputs, for inhomogeneous deformation input fields, the monotonicity filters should be turned off by setting $C_{\rm mono}=C_{\rm vol}=0$.
\end{itemize}

\begin{table}[!t]
\setlength{\tabcolsep}{4.0pt} 
\def\arraystretch{1.25}
\centering
{\footnotesize
 \begin{tabular}{|c|c|c|c|c|c|c|c|}
 \hline
 \multirow{2}{*}{Parameter} & \multirow{2}{*}{Description} & 1st-func & 1st-func & 1st-func & 2nd-func & 2nd-func &  \multirow{2}{*}{Noise study} \\
 & & low density & medium density & high density & homogeneous & inhomogeneous & \\
 \hline
 $\alpha$ & Weight in Eq.~\eqref{obj} & 2.55 & 2.55 & 2.2 & 1.7 & 1.0 & 1.0/1.7/3.0 \\
 \hline
 $n_{\rm ten}$ & Load steps in tension & 33 & 33  & 40  & 13 & \multirow{2}{*}{7} & 9\\
 $n_{\rm comp}$ & Load steps in compression  &  100 & 100  & 100  & 17 &  & 16 \\
 \hline
 \multirow{2}{*}{$(K_1,K_2)_{\rm c}$} & Loss of ellipticity check & \multicolumn{6}{c|}{\multirow{2}{*}{(-0.15, 0.15), (-0.25, 0.25), (-0.35, 0.35), (-0.45, 0.45), (-0.55, 0.55)}} \\
 & points in compression ($K_3 = -1$) & \multicolumn{6}{c|}{} \\
 \hline
 \multirow{2}{*}{$(K_1,K_2)_{\rm s}$} & Loss of ellipticity check  & \multicolumn{6}{c|}{\multirow{2}{*}{(0, 0.51)}} \\
 & points in shear ($K_3 = 0$) & \multicolumn{6}{c|}{} \\
 \hline
 $C_{\rm mono}$ & Stress monotonicity filter &  \multicolumn{4}{c|}{1 (applied)} & \multicolumn{2}{c|}{0 (not applied)} \\
 \hline
 $C_{\rm vol}$ & Vol. strain monotonicity filter & \multicolumn{4}{c|}{1 (applied)} & \multicolumn{2}{c|}{0 (not applied)} \\
 \hline
 \end{tabular}}
\caption{VFM-related parameters used in the applications of the 1st-func of the VFM-GA framework in Section~\ref{sec:1st} for low, medium and high density Poron XRD foams; the applications of the VFM-GA 2nd-func in Section~\ref{sec:2nd} for both homogeneous and inhomogeneous deformation input fields; and the noise study of Section~\ref{sec:noise}.} 
\label{tab:VFM_parameters}
\end{table}

\subsection{GA hyperparameters}\label{app:hyper para}
Table~\ref{tab:GA_hyperparameters} gives the GA hyperparameters used in the applications of the 1st-func of the VFM-GA framework in Section \ref{sec:1st}, in the applications of the VFM-GA 2nd-func in Section~\ref{sec:2nd}, and in the noise study of Section~\ref{sec:noise}. Most of the GA hyperparameters are consistent throughout this work. The exceptions are $n_{\rm gen}$ and $n_{\rm t}$, which are lowered for the 2nd-func applications compared to the 1st-func applications to reduce computational cost, and the parameter $m_{\rm rate}$, which is decreased in the 2nd-func applications due to the empirically observed change in the GA's sensitivity to mutation between the two functionalities.  
Descriptions of the GA hyperparameters may be found in Section~\ref{sec:GA} and are recapped below. 
\begin{itemize}
    \item {$n_{\rm param}$}: One must set $n_{\rm param}$ to the number of material parameters in the designated constitutive model. 
    \item {$n_{\rm b}$}: The hyperparameter $n_{\rm b}$ is the number of bits used in the discretization of parametric space. It is interrelated with the choices of the prescribed ranges for the material parameters discussed in Appendix~\ref{app:mat para}. Using a higher value of $n_{\rm b}$ leads to higher precision in the calibrated material parameters but at a higher computational cost. In our experience, we find that $n_{\rm b} = 4$ works well if the ranges between the maximum and minimum limits are within two orders of magnitude. If an application of the VFM-GA framework results in calibrated material parameters that fall within their respective prescribed ranges but do not perform satisfactorily, one should increase $n_{\rm b}$ to increase the precision and rerun the code package before judging the effectiveness of a constitutive model.
    \item {$n_{\rm gen}$}: The hyperparameter $n_{\rm gen}$ is the number of generations for each independent population. Based on our experience, one should set $n_{\rm gen} \ge 70$ to prevent premature termination of the evolution cycle; however, we have observed that taking $n_{\rm gen} \ge 150$ results in  diminishing returns. 
    \item {$n_{\rm t}$}: The hyperparameter $n_{\rm t}$ is the number of candidate material parameter sets in each independent population. As mentioned in Section \ref{sec:strat}, it is recommended to choose $n_{\rm t} \le 1000$. However, if sufficient computational resources are available, one can choose a higher value of $n_{\rm t}$ but should not exceed 2000, which is the point at which we have observed diminishing returns.
    \item {$n_{\rm pop}$}: The hyperparameter $n_{\rm pop}$ is the number of independent initial populations. As mentioned in Section \ref{sec:strat}, it is recommended to take $n_{\rm pop} \ge 400$, but diminishing returns are expected for $n_{\rm pop} \ge 1000$. 
    \item{$T_{\rm max}$ and $T_{\rm min}$}: These are the two hyperparameters in \eqref{eq:boltz} defining the maximum and minimum quasi-tem\-per\-a\-tures used in Boltzmann Selection in the GA (Section~\ref{sec:boltz}). It is recommended to set $T_{\rm min}$ to a value close to 0, and $T_{\rm max}$ should be on the same order as the minimum objective function value across the initial populations. Here, we take $T_{\rm max} = 60$ and $T_{\rm min} = 0.5$ across the applications in the present work. 
    \item {$x_{\rm rate}$ and $x_{\rm con}$}: These are the two hyperparameters controlling the probability of crossover in the GA at the set level and parameter level, respectively (Section~\ref{sec:crossover}). Based on our experience, we recommend setting $x_{\rm rate} = 0.84$ and $x_{\rm con} = 0.9$. 
    \item {$m_{\rm rate}$ and $m_{\rm con}$}: These are the two hyperparameters controlling the probability of mutation in the GA at the set level and parameter level, respectively (Section~\ref{sec:mutation}). Based on our experience, it is recommended to set $m_{\rm con} = 0.2$, and $m_{\rm rate}$ should be set to 0.7 when using the 1st-func and 0.03 when using the 2nd-func to accommodate the increased sensitivity to mutation observed for the 2nd-func.
\end{itemize}

\begin{table}[!t]
\setlength{\tabcolsep}{3.92pt} 
\def\arraystretch{1.25}
\centering{
\footnotesize
 \begin{tabular}{|c|c|c|c|c|c|c|c|}
 \hline
 \multirow{2}{*}{Parameter} & \multirow{2}{*}{Description} & 1st-func & 1st-func & 1st-func & 2nd-func & 2nd-func & Noise \\
 & & low density & medium density &  high density & homogeneous & inhomogeneous & study \\
 \hline
 $n_{\rm param}$ & Number of material parameters in $\boldsymbol{\theta}$ & \multicolumn{6}{c|}{14} \\ 
 \hline
 $n_{\rm b}$ & Number of bits & \multicolumn{6}{c|}{4} \\
 \hline
 $n_{\rm gen}$ & Number of generations per population & \multicolumn{3}{c|}{100} & \multicolumn{3}{c|}{80}\\
 \hline
 $n_{\rm t}$ & Number of candidates in a population & \multicolumn{3}{c|}{500} & \multicolumn{3}{c|}{300}  \\
 \hline
 $n_{\rm pop}$ & Number of independent populations & \multicolumn{6}{c|}{500} \\
 \hline
 $T_{\rm max}$ & Max quasi-temperature & \multicolumn{6}{c|}{60} \\
 \hline
 $T_{\rm min}$ & Min quasi-temperature & \multicolumn{6}{c|}{0.5}  \\
 \hline
 $x_{\rm rate}$ & Probability of set-level crossover & \multicolumn{6}{c|}{0.84} \\
 \hline
 $x_{\rm con}$ & Probability of parameter-level crossover & \multicolumn{6}{c|}{0.9} \\
 \hline
 $m_{\rm rate}$ & Probability of set-level mutation & \multicolumn{3}{c|}{0.7} & \multicolumn{3}{c|}{0.03}  \\
 \hline
 $m_{\rm con}$ & Probability of parameter-level mutation & \multicolumn{6}{c|}{0.2} \\
 \hline
 \end{tabular}}
\caption{GA hyperparameters used in the applications of the 1st-func of the VFM-GA framework in Section~\ref{sec:1st} for low, medium and high density Poron XRD foams; the applications of the VFM-GA 2nd-func in Section~\ref{sec:2nd} for both homogeneous and inhomogeneous deformation input fields; and the noise study of Section~\ref{sec:noise}.} 
\label{tab:GA_hyperparameters}
\end{table}

\end{document}